\documentclass[final]{elsarticle}
\usepackage[includeheadfoot, margin = 2.5 cm]{geometry}

\usepackage{url}
\usepackage{fancyvrb}
\usepackage{fvextra}
\usepackage{inconsolata}
\usepackage{enumitem}
\usepackage{mathtools}
\usepackage{mdframed}
\newmdenv[
  topline=true,
  bottomline=true,
  leftline=true,
  rightline=true,
  skipabove=1em,
  skipbelow=1em
]{myframe}
\usepackage{threeparttable}
\usepackage{braket}
\usepackage[normalem]{ulem}

\newcommand{\U}{$U$}
\newcommand{\ve}{\varepsilon}
\usepackage{amssymb}
\usepackage{amsmath}

\biboptions{sort&compress}
\usepackage[
  colorlinks=true,
  linkcolor=blue,
  citecolor=blue,
  urlcolor=blue,
  linktocpage=true
]{hyperref}

\usepackage{fontawesome5}

\journal{Computer Physics Communications}

\begin{document}

\begin{frontmatter}



\title{\texttt{FlaSR}: A Mathematica Package for the Automatic Generation of $U$-spin Sum Rules}

\author[caltech]{Margarita Gavrilova}
\ead{mgavr@caltech.edu}
\author[cornell]{Livia Kong}
\ead{lk386@cornell.edu}

\affiliation[caltech]{organization={Walter Burke Institute for Theoretical Physics, California Institute of Technology},
             city={Pasadena},
             postcode={91125},
             state={CA},
             country={USA}}

\affiliation[cornell]{organization={Department of Physics, LEPP, Cornell University},
             city={Ithaca},
             postcode={14853},
             state={NY},
             country={USA}}

\begin{abstract}
We present \texttt{FlaSR}, a Mathematica package for the automatic generation of higher-order amplitude and amplitude-squared $SU(2)$ flavor sum rules, with a particular emphasis on $U$-spin. The package is based on recent insights into the mathematical structure of flavor sum rules. Given a system of symmetry-related processes defined by a list of particle multiplets in the initial and final states, together with the transformation properties of the symmetry-limit Hamiltonian, the package generates the full list of processes and outputs amplitude and amplitude-squared sum rules for the system. For amplitude sum rules, the package generates the complete set at all orders in symmetry breaking for which such relations exist. For amplitude-squared sum rules, which translate directly into sum rules among physical observables, the current version of the package is not guaranteed to generate the complete set. We illustrate the scope of the package with a range of representative examples.

\end{abstract}




\end{frontmatter}

\clearpage
\section*{Program Summary}

\noindent
\textit{Program Title:} \texttt{FlaSR}

\medskip
\noindent
\textit{Developer's repository link:}
\url{https://github.com/Flavor-Sum-Rules/FlaSR}

\medskip
\noindent
\textit{Licensing provisions:} MIT license (MIT)

\medskip
\noindent
\textit{Programming language:} Wolfram Language (Mathematica), Python

\medskip
\noindent
\textit{Supplementary material:} Example tutorial notebook, documentation

\medskip
\noindent
\textit{Nature of problem:} Reliable theoretical predictions for fully hadronic weak decays are challenging due to the non-perturbative dynamics of quantum chromodynamics (QCD). Approximate flavor symmetries, in particular $U$-spin, impose model-independent relations called sum rules among symmetry-related amplitudes and decay rates. These relations are especially powerful when derived beyond the exact symmetry limit since higher-order sum rules yield more stringent theoretical predictions. In practice, however, deriving such relations with the standard Clebsch--Gordan decomposition method becomes a cumbersome bookkeeping and classification task. Large systems involve many decay channels, arbitrary external multiplets, and higher-order spurion insertions. The standard derivation must be repeated separately for each system and becomes inefficient for large multibody applications and broad surveys of higher-order amplitude and amplitude-squared sum rules. Furthermore, the resulting relations appear as an arbitrary null-space basis of a Clebsch--Gordan coefficient matrix, which obscures their symmetry structure and organization by order in the symmetry breaking.\\

\medskip
\noindent
\textit{Solution method:} The \texttt{FlaSR} package implements the systematic method for deriving $SU(2)$ flavor sum rules developed in Refs.~\cite{gavrilova2022structure,Gavrilova:2026ryc} and upcoming publications. The method relies on a series of theorems that characterize the structure of $U$-spin amplitude sum rules and allow their determination without performing a Clebsch--Gordan decomposition. Amplitude-squared sum rules are then derived from the amplitude sum rules following the procedure described in the text. The package generates the full set of processes and outputs amplitude and amplitude-squared sum rules for $U$-spin systems. The program output is guaranteed to contain the complete set of amplitude sum rules at each order in flavor-symmetry breaking at which such relations exist, but the completeness is not guaranteed at the amplitude-squared level. The user interface is written in Mathematica, with computationally intensive routines implemented in Python.\\

\medskip
\noindent
\textit{Additional comments including restrictions and unusual features:} The program requires a Python installation; instructions for configuring Python for use with Mathematica are provided online.\\

\clearpage

\tableofcontents
\clearpage



\section{Introduction}
\label{sec:intro}

Fully hadronic weak decays remain theoretically difficult due to the presence of non-perturbative QCD dynamics. The approximate $SU(3)_F$ flavor symmetry of the light quarks $u$, $d$, and $s$ provides a model-independent way to study hadronic weak decays and other hadronic processes. In this work we focus on $U$-spin, the $SU(2)$ subgroup of $SU(3)_F$ that relates the $d$ and $s$ quarks. While symmetry alone does not allow one to calculate amplitudes or rates from first principles, it can imply linear relations among them. We refer to linear relations among amplitudes as amplitude sum rules, and to the corresponding relations among decay rates as rate sum rules. In many cases, these symmetry-based relations remain as state-of-the-art theoretical predictions.

$U$-spin is, however, known to be badly broken. At the fundamental level the breaking originates from the quark mass difference $\Delta m = m_s - m_d$, and thus the breaking effects are expected to be proportional to $\Delta m$. It is common to introduce a small parameter controlling the breaking as $\varepsilon = \Delta m/\Lambda$, where $\Lambda$ is a characteristic scale commonly taken to be of order $\Lambda_{\text{QCD}}$, which implies $\varepsilon \sim 20\text{--}30\%$. In what follows we adopt this crude estimate and assume that one can indeed define a small parameter $\varepsilon \propto \Delta m$. The $U$-spin breaking can then formally be incorporated in the flavor symmetry analysis through insertions of the mass spurion. By systematically expanding in powers of the symmetry-breaking spurion, one can derive sum rules that hold beyond the exact symmetry limit and are therefore expected to be satisfied with higher precision.

Such higher-order sum rules, when confronted with experimental data, provide tests of the flavor-symmetry framework. Deviations from these relations can probe the pattern of flavor-symmetry breaking, provide insights into non-perturbative QCD dynamics, and potentially offer probes of physics beyond the Standard Model. The increasing need for a more comprehensive understanding of the implications of flavor symmetries is further highlighted by recent experimental advances in measurements of hadronic decays ~\cite{LHCb:2019hro, LHCb:2019oke, LHCb:2022lry, LHCb:2022lzp, LHCb:2023mwc, LHCb:2023rae, Belle-II:2023vra, Belle:2023bzn, Belle:2023str, LHCb:2023ngz, LHCb:2023tma, CMS:2024hsv, LHCb:2024rkp, LHCb:2024vhs, LHCb:2024hfo, Belle-II:2024vfw, Belle-II:2024xtf, Belle:2024dhj, Belle-II:2024frs, Belle:2024ymt, LHCb:2024yzj, LHCb:2024iuo, LHCb:2024jpt, Belle-II:2025xvc, LHCb:2025zgk, LHCb:2025qcs, LHCb:2025uxu, LHCb:2025ray, Belle-II:2025zqj, Belle:2025voy, LHCb:2025drb, LHCb:2025hul, LHCb:2025zvw, LHCb:2025ozp, LHCb:2025ftm, LHCb:2025lwm, LHCb:2025nys, LHCb:2026jgu}, including multibody decays, as well as future prospects~\cite{Belle-II:2018jsg, Cerri:2018ypt, LHCb:2018roe} and the ongoing discussion of whether current data shows indications of larger-than-expected flavor-symmetry breaking~\cite{Schacht:2022kuj,Huber:2021cgk, Bhattacharya:2022akr, Berthiaume:2023kmp, Bhattacharya:2025wcq, BurgosMarcos:2025xja, Bhattacharya:2026stc}.

There exists an extensive literature on symmetry-limit flavor sum rules~\cite{Shmushkevich:1955, DushinShmushkevich:1956, MacfarlaneMukundaSudarshan:1964, PinskiMacfarlaneSudarshan:1965, LipkinPeshkin1972Inclusive, Kyriakopoulos1974IsospinRelations, Altarelli:1974sc, Kingsley:1975fe, Zeppenfeld:1980ex, Savage:1991wu, Chau:1991gx, Chau:1993ec, Buccella:1994nf, Gronau:1995hm, Fleischer:1999pa, Gronau:2000md, Gronau:2000zy, Grossman:2003qp, Buras:2004ub, Grossman:2006jg, Jung:2009pb, Pirtskhalava:2011va, Feldmann:2012js, Franco:2012ck, Brod:2012ud, Cheng:2012xb, Atwood:2012ac, Hiller:2012xm, Grossman:2013lya, Jung:2014jfa, Muller:2015lua, Muller:2015rna, Ligeti:2015yma, Grossman:2018ptn, Wang:2019dls, Buccella:2019kpn, Wang:2022kwe, Wang:2023pnb, Luo:2023vbx, Wang:2024nxb, Fu:2025wwx, Zhang:2025jnw, Wen:2025ibn, Liu:2025hbf}. Examples of second-order flavor sum rules were pointed out in Refs.~\cite{grossman2013su3, Brod:2012ud, Wen:2025ibn}, second-order non-linear relations were given in Refs.~\cite{Gronau:2013xba, Gronau:2015rda}, and some general results for higher-order amplitude sum rules were obtained in Ref.~\cite{Hassan:2022ucn}. A systematic analysis of higher-order amplitude sum rules was carried out in Refs.~\cite{gavrilova2022structure, gavrilova2024direct}. This analysis revealed the rich underlying structure of amplitude sum rules and developed a simple and universal formalism allowing the derivation of $U$-spin amplitude sum rules for arbitrary systems of processes related by $U$-spin and to arbitrary order in symmetry breaking without any reliance on explicit Clebsch--Gordan decomposition. Beyond simplifying the derivation, this formalism makes the organization of the sum rules transparent. For a given system, it determines the number and form of the independent relations at each order in the symmetry breaking, identifies the highest order at which any sum rule can exist, and reveals a universal structure shared by systems built from the same collection of $SU(2)$ irreps, independently of whether those irreps appear in the initial state, final state, or Hamiltonian. This structural understanding also provides the starting point for a systematic analysis of relations among rates, including the effects of channel-dependent phase space. Some general results for second-order rate sum rules were recently presented in Ref.~\cite{Gavrilova:2026ryc}.

Here we present \texttt{FlaSR}\footnote{Pronounced ``flaser"---that is, ``laser" with an ``f" in front.}, a Mathematica package for the automatic generation of higher-order $U$-spin sum rules. The package is based on Refs.~\cite{gavrilova2022structure, Gavrilova:2026ryc} and thus does not perform an explicit Clebsch--Gordan decomposition, but instead relies on the theorems derived in those references and on the algorithmic rules for constructing sum rules developed therein. The \verb|FlaSR| package takes as input lists of the particle multiplets in the initial and final states along with the irreducible representation (irrep) of the symmetry-limit Hamiltonian, which together define a system of processes related by the symmetry. The output consists of amplitude sum rules as well as sum rules between the squared moduli of amplitudes for the system. The amplitude sum rules are derived to all orders in the symmetry breaking for which they exist. The rate sum rules are derived from the amplitude sum rules according to a procedure we outline in this paper. This procedure is not guaranteed to produce a full set of rate sum rules. Thus, we emphasize that additional rate sum rules beyond those generated by the package may exist. We note that, under the assumptions of Ref.~\cite{Gavrilova:2026ryc}, second-order amplitude-squared sum rules imply second-order sum rules for the corresponding differential or integrated rates and are therefore of particular phenomenological interest. An extension of this analysis to higher orders in symmetry breaking will be presented in forthcoming work.

Although \verb|FlaSR| is primarily intended for applications to weak hadronic decays related by $U$-spin, the underlying results are general group-theoretical statements that apply to any $SU(2)$ system satisfying the assumptions summarized in Section~\ref{sec:scope}. In particular, the particles in the external states transform in irreducible representations of $SU(2)$, the symmetry-limit Hamiltonian transforms in an irreducible representation, the system of amplitudes is complete, and the symmetry breaking is described perturbatively by a triplet spurion. The applicability of the formalism to other $SU(2)$ subgroups, such as isospin and $V$-spin, and to processes other than weak decays depends on whether these conditions are satisfied.

This paper is organized as follows. In Section~\ref{sec:theory} we briefly review the definitions and main results of Refs.~\cite{gavrilova2022structure, Gavrilova:2026ryc} necessary for working with \texttt{FlaSR}. In Section~\ref{sec:instructions} we provide instructions for installing and running the package. In Section~\ref{sec:examples} we collect representative examples of its use. In Section~\ref{sec:conclusion} we conclude.

\section{Review of theory and definitions}
\label{sec:theory}

The \texttt{FlaSR} package is based on the general group-theoretical framework for $SU(2)$ flavor sum rules developed in Refs.~\cite{gavrilova2022structure,Gavrilova:2026ryc} and on forthcoming work on amplitude-squared $SU(2)$ flavor sum rules. These works leverage the mathematical structure of $SU(2)$ flavor sum rules to formulate a systematic procedure for deriving them at arbitrary orders in symmetry breaking without performing an explicit Clebsch--Gordan decomposition. The algorithms implemented in \texttt{FlaSR} follow directly from these results.

In this section, we briefly review the definitions and key results needed to understand and use the \verb|FlaSR| package. We refer the reader to Refs.~\cite{gavrilova2022structure,gavrilova2024direct,Gavrilova:2026ryc} for further details and proofs. In what follows, we use weak decays related by $U$-spin as the main language for presenting the formalism and the examples implemented in \texttt{FlaSR}. This reflects the original motivation of the work in Refs.~\cite{gavrilova2022structure,gavrilova2024direct,Gavrilova:2026ryc}, which was driven by phenomenological applications in which higher-order $U$-spin sum rules play an important role. At its core, however, the formalism is simply based on $SU(2)$ symmetry. For this reason, the methods implemented in \texttt{FlaSR} can also be applied more generally to other $SU(2)$ subgroups of flavor symmetry, namely isospin and $V$-spin, and to processes that go beyond weak decays. We comment on such applications in Section~\ref{sec:scope}, where we also summarize the assumptions underlying the algorithm and the scope of the results.

\subsection{Definitions and notation}
\label{sec:defs}

\paragraph{\U-spin system} The crux of flavor sum rules lies in the approximate flavor symmetry respected by QCD. In this work, we focus on \U-spin, the $SU(2)$ symmetry between $d$ and $s$ quarks. Under \U-spin, $d$ and $s$ quarks form doublets and transform as:
\begin{equation}
    \begin{bmatrix}
        d \\[4pt]
        s
    \end{bmatrix}
    =
    \begin{bmatrix}
        \ket{\frac{1}{2},+\frac{1}{2}} \\[4pt]
        \ket{\frac{1}{2},-\frac{1}{2}}
    \end{bmatrix}
    ,\quad
    \begin{bmatrix}
        \overline{s} \\[4pt]
        -\overline{d}
    \end{bmatrix}
    =
    \begin{bmatrix}
        |\frac{1}{2},+\frac{1}{2}\rangle \\[4pt]
        |\frac{1}{2},-\frac{1}{2}\rangle
    \end{bmatrix}
\,.\end{equation}
We indicate the quantum numbers (QNs) as $|u,m\rangle$, where $u$ is the total \U-spin and $m$ is the third component of \U-spin. Note that throughout this paper, we label irreps by their total $SU(2)$ angular momentum $u$, not by the number of states in the multiplet, $2u+1$.

A \U-spin system is the full set of amplitudes related by $U$-spin symmetry. To define a \U-spin system, one must specify the representations in the initial and final states, as well as the representation of the symmetry-limit effective Hamiltonian, which we denote as $\mathcal{H}_\text{eff}^{(0)}$. We assume that all particles in the external states and the symmetry-limit Hamiltonian transform as irreps of $U$-spin. In particular we take 
\begin{equation}\label{eq:Heff0}
    \mathcal{H}_{\mathrm{eff}}^{(0)} = \sum_{m} f_{u,m} H_m^u\,,
\end{equation}
where $H_m^u$ are operators with total $U$-spin $u$ and third component of $U$-spin $m$, while $f_{u,m}$, in the case of weak decays, encode the corresponding CKM factors.

The physical amplitudes of the system are given by
\begin{equation}
    \mathcal{A}_i \equiv \langle \text{out} \lvert \mathcal{H}_{\text{eff}} \rvert \text{in} \rangle_i \,,
\end{equation}
where $i$ is an index that enumerates the amplitudes, and $\mathcal{H}_\text{eff}$ is, in general, the effective Hamiltonian to all orders in flavor-symmetry breaking. By our assumption of pure irreps, there is a single value of $u$ in $\mathcal{H}_{\mathrm{eff}}^{(0)}$, and we therefore can define the CKM-free amplitudes as
\begin{equation}
    A_i \equiv \frac{\mathcal{A}_i}{f_{u,m}} \,.
\end{equation}
After identifying the irreps describing the system and defining the CKM-free amplitudes, the task of finding sum rules becomes a purely group-theoretical problem. In practice, we thus derive sum rules for CKM-free amplitudes and restore the CKM factors only at the very end. With this in mind, ``$A$-type amplitudes'' should henceforth be understood to refer to CKM-free amplitudes unless CKM factors are shown explicitly.

\paragraph{The running example} To illustrate the ideas throughout this paper, we use the system $D^0 \to P^+ P^-$ as a running example. Its external states are described by the following $U$-spin multiplets:
\begin{equation}
D^0 = |c\overline{u}\rangle\,, \quad
P^+ =
\begin{bmatrix}
    K^+ \\[2pt]
    \pi^+
\end{bmatrix}
=
\begin{bmatrix}
    |u\overline{s}\rangle \\[2pt]
    -|u\overline{d}\rangle
\end{bmatrix}, \quad
P^- =
\begin{bmatrix}
    \pi^- \\[2pt]
    K^-
\end{bmatrix}
=
\begin{bmatrix}
    |d\overline{u}\rangle \\[2pt]
    |s\overline{u}\rangle
\end{bmatrix}
\label{eq:DPPmultiplet}
\,.
\end{equation}
In the $U$-spin limit, the effective Hamiltonian $\mathcal{H}_\text{eff}^{(0)}$ for this system is approximated as a triplet. Accordingly, the initial state has $u_{D^0}=0$, the final-state multiplets have $u_{P^+}=u_{P^-}=1/2$, and the Hamiltonian has $u_H=1$, where the subscripts label the physical multiplets corresponding to the $U$-spin irreps. We present the full \texttt{FlaSR} output for this example in Section~\ref{sec:D to PP}, where we also justify the approximation in the Hamiltonian.\\

We now describe several conventions for labeling amplitudes, together with a few remarks on their implementation in the \verb|FlaSR| package. We introduce the following types of labels:
\begin{itemize}
    \item physical process labels,
    \item $m$-QN labels,
    \item $n$-tuple labels.
\end{itemize}

\paragraph{Physical processes} For physical systems, where the physical particle states of the multiplets are specified, the physical processes themselves can be used to label amplitudes. For example, in the $D^0 \to P^+ P^-$ system, one of the decays is $D^0 \to \pi^+ K^-$. We label the corresponding amplitude as $A(D^0 \to \pi^+ K^-)$.

\paragraph{Quantum numbers} Every amplitude in a \U-spin system can be labeled by listing the $m$-QNs of the participating components of the multiplets. For the $D^0 \to \pi^+ K^-$ decay, $D^0 = |0,0\rangle$, $\pi^+ = |\frac{1}{2},-\frac{1}{2}\rangle$, $K^- = |\frac{1}{2},-\frac{1}{2}\rangle$, and the process is realized by the $H_{-1}^1$ operator. Thus this amplitude can be written in the $m$-QN notation as $A(0 \overset{-1}{\to} -\frac{1}{2} -\frac{1}{2})$.

\paragraph{$n$-tuples} The $n$-tuple notation is rooted in the observation that any \U-spin system can be built from doublets, since every higher $SU(2)$ irrep can be obtained from tensor products of the fundamental representation. We use $n$ to denote the minimal number of such would-be doublets needed to construct all irreps in the system (singlet irreps are excluded from this counting, since their $m$-QN is fixed across all amplitudes of the system). To each component of a \U-spin multiplet with irrep $u$, we assign a string of $2u$ plus and minus signs, with `$-$' and `$+$' corresponding to the $-1/2$ and $+1/2$ components of a would-be doublet, respectively. For a component of $u$ with $m$-QN $m$, the associated string is
\begin{equation}
\underbrace{--\cdots-}_{u-m}
\underbrace{++\cdots+}_{u+m} \,,
\end{equation}
so that the sum of the signs reproduces $m$; by convention, all minus signs are written before all plus signs. The $n$-tuple of an amplitude is obtained by ordering the irreps and joining the strings assigned to the participating components. For final-state irreps, we use the string associated with $m$, while for irreps in the initial state and in the Hamiltonian, we use the string associated with $-m$. With this convention, valid amplitudes correspond to $n$-tuples with equal numbers of `$-$' and `$+$' signs.

Consider the $D^0 \to P^+ P^-$ system. For this system, $n=4$. We fix the order of the non-singlet irreps as $u_{P^+},\, u_{P^-},\, u_H$. The components entering the amplitude $D^0 \to \pi^+ K^-$ have $m$-QNs $m_{P^+}=-1/2$, $m_{P^-}=-1/2$, and $m_H=-1$. We therefore label this amplitude as $A_{(-,-,++)}$.\\

Finally, we note that in \verb|FlaSR|, when amplitudes are labeled by physical processes (or by $m$-QNs), the components of the multiplets (or the corresponding $m$-QNs) are written in the order specified by the user input. In contrast, for $n$-tuple labels, \verb|FlaSR| uses an internal ordering convention for the irreps rather than the user-provided order: internally, the irreps are ordered from smallest to largest, with ties resolved according to their order in the user input.

\subsection{Amplitude sum rules}

Given a \U-spin system, our first goal is to derive a complete set of linearly independent amplitude sum rules among CKM-free amplitudes that hold up to a fixed order $b$ in the symmetry breaking. Such sum rules take the form
\begin{equation}
    \sum_i \alpha_i A_i = \mathcal{O}(\varepsilon^{b+1}) \,,
\end{equation}
where the $\alpha_i$ are numerical coefficients. We say that a relation holds ``up to order $b$,'' or simply ``at order $b$,'' if it is violated by corrections of order $b+1$.

\paragraph{Amplitude pairs} When deriving sum rules, it is convenient to work with \U-spin amplitude pairs. A \U-spin amplitude pair consists of an amplitude $A_i$ and its \U-spin conjugate $\overline{A}_i$\,\footnote{Here the bar denotes \U-spin conjugation, not CP conjugation.}. The amplitude $\overline{A}_i$ is obtained from $A_i$ by interchanging all $d$ and $s$ quarks or, equivalently, by reversing the signs of all $m$-QNs. For example, in the $D^0 \to P^+ P^-$ system, the amplitudes
\begin{align}\label{eq:pair_example}
    A(D^0 \to \pi^+ K^-) &\equiv A(0 \overset{-1}{\to} -\tfrac{1}{2}\, -\tfrac{1}{2}) \equiv A_{(-,-,++)} \,, \nonumber\\
    A(D^0 \to K^+ \pi^-) &\equiv A(0 \overset{+1}{\to} +\tfrac{1}{2}\, +\tfrac{1}{2}) \equiv A_{(+,+,--)}
\end{align}
form a \U-spin amplitude pair.

For each amplitude pair, we can perform a basis change from the basis $(A_i,\,\overline{A}_i)$ to the basis $(a_i,\,s_i)$. We define $a$($s$)-type amplitudes as\footnote{Our convention differs slightly from that of Ref.~\cite{gavrilova2022structure}: the additional factor $(-1)^{q_i}$ appearing there is not included explicitly in the definitions of $a_i$ and $s_i$. In \texttt{FlaSR}, these factors are instead taken into account when writing the corresponding sum rules.}
\begin{equation}
    a_i \equiv A_i - (-1)^p \overline{A}_i\,,\qquad
    s_i \equiv A_i + (-1)^p \overline{A}_i\,,
    \label{eq:asdef}
\end{equation}
where $p$ is a system-dependent integer factor discussed in Appendix~C of Ref.~\cite{gavrilova2022structure} and calculated automatically by \texttt{FlaSR}. The $p$-factor is chosen so that $a$-type amplitudes are antisymmetric under \U-spin conjugation, while $s$-type amplitudes are symmetric. With this basis choice, $a$-type amplitudes contain only odd powers of the symmetry-breaking parameter, while $s$-type amplitudes contain only even powers. As a result, the $a$- and $s$-type sectors decouple, and sum rules can be written separately in terms of $a$-type or $s$-type amplitudes. Moreover, an $a$-type ($s$-type) sum rule that holds through odd (even) order $b$ automatically holds through order $b+1$. We also note that the highest order at which an $a$-type ($s$-type) sum rule can hold is always even (odd). \texttt{FlaSR} leverages the simplifications offered by the $(a_i,\,s_i)$ basis to derive sum rules in this basis. The sum rules can then be rewritten in terms of $(A_i,\, \overline{A}_i)$ using the definitions in Eq.~\eqref{eq:asdef}.

The $a$- and $s$-type amplitudes are indexed by the underlying amplitude pair. By convention, the label is chosen to match that of the member of the pair whose $n$-tuple begins with a minus sign. For the $D^0\to P^+ P^-$ example above, $(-1)^p=1$, and
\begin{equation}
    a_{(-,-,++)} \equiv A_{(-,-,++)} - A_{(+,+,--)}\,,
    \qquad
    s_{(-,-,++)} \equiv A_{(-,-,++)} + A_{(+,+,--)}\,.
\end{equation}
If all irreps in the system are integers, then one amplitude---the one for which the $m$-QNs of all irreps are equal to zero---is self-conjugate. For that amplitude, one of $a_i$ or $s_i$ is identically zero and the other is identically $2A_i$; this is a definition, not a sum rule, and \texttt{FlaSR} handles it automatically.

\paragraph{Coordinate notation} The all-order sum rule construction is most transparent in coordinate notation, where each amplitude pair is mapped to a point on a $d$-dimensional lattice,
\begin{equation}
    d \equiv n/2 - 1\,,
\end{equation}
where $n$ is the number of would-be doublets. For simplicity in explaining the lattice construction, we assume that the system is described by $r$ non-singlet irreps and that at least one of them is a doublet. \texttt{FlaSR} considers systems without doublets in its implementation as well; the treatment of these cases is discussed in Section~V.B, step~2.3, of Ref.~\cite{gavrilova2022structure}, and the additional symmetrization step used there is explained in Section~IV.B. To obtain the coordinate of an amplitude pair, we first order the irreps so that a doublet comes first and label them as $0,1,\ldots,r-1$. We then label the strings in the $n$-tuple by the same numbers as their corresponding irreps. The coordinate of the amplitude pair is obtained by listing the positions of all minus signs in the $n$-tuple, omitting the first minus sign. Recall that, by convention, amplitude pairs are labeled by an $n$-tuple that begins with a minus sign. For the running example, $d=1$, and for the pair in Eq.~\eqref{eq:pair_example} we have
\begin{equation}
    (\underset{0}{-},\underset{1}{-},\underset{2}{++}) \longrightarrow (1) \,,
\end{equation}
where $(1)$ on the right-hand side is a coordinate in a one-dimensional lattice. When $d > 1$ all permutations of coordinates correspond to the same amplitude pair.

\paragraph{Reading off sum rules} The sum rules are then read off from the lattice. To derive sum rules that hold through order $b$, we form weighted sums over $b$-dimensional subspaces of the lattice. The corresponding weights are discussed in Section~IV.C.2 and Eq.~(4.34) of Ref.~\cite{gavrilova2022structure}. For even (odd) $b$, these sums yield $a$-type ($s$-type) sum rules. Since the largest subspace is the full lattice, the highest order at which the system admits sum rules is
\begin{equation}
    b_{\max} = d \equiv n/2 - 1 \,.
\end{equation}

\subsection{Amplitude-squared sum rules}

Having obtained amplitude sum rules, we can proceed to derive amplitude-squared sum rules. Amplitude-squared sum rules that hold at order $b$ in the symmetry breaking are linear relations of the form
\begin{equation}
    \sum_i \gamma_i |A_i|^2 = \mathcal{O}(\varepsilon^{b+1}) \,,
\end{equation}
where the $\gamma_i$ are numerical coefficients. Such sum rules are of particular phenomenological interest. As shown in Ref.~\cite{Gavrilova:2026ryc}, amplitude-squared sum rules that hold up to corrections of $\mathcal{O}(\varepsilon^2)$ straightforwardly imply sum rules among CKM-normalized integrated decay rates for arbitrary multibody decays upon the replacement $|A_i|^2 \to \Gamma_i$. They can therefore be directly tested in experiment. An extension of this analysis to arbitrary orders is deferred to future work. For the purposes of this work, we proceed under the conjecture that this correspondence extends to arbitrary orders, so that any amplitude-squared sum rule valid up to corrections of $\mathcal{O}(\varepsilon^{b+1})$ implies the corresponding sum rule among CKM-normalized decay rates with corrections of the same order.

In direct analogy with the $a$($s$)-type amplitude basis, it is useful to define a basis of $\Delta$($\Sigma$)-type amplitudes-squared:
\begin{equation}
    \Delta_i \equiv |A_i|^2 - |\overline{A}_i|^2\,,\qquad
    \Sigma_i \equiv |A_i|^2 + |\overline{A}_i|^2 \,.
    \label{eq:deltasigmadef}
\end{equation}
Unlike in Eq.~\eqref{eq:asdef}, no $p$-factors appear here, so the relative signs in the definitions of $\Delta_i$ and $\Sigma_i$ are fixed. By construction, $\Delta_i$ are anti-symmetric under $U$-spin conjugation, while $\Sigma_i$ are symmetric. In this basis, amplitude-squared sum rules exhibit a decoupling analogous to that of amplitude sum rules in the $a$($s$)-type basis: amplitude-squared sum rules can be written entirely in terms of either $\Delta$-type or $\Sigma$-type amplitudes-squared. Moreover, a $\Delta$-type ($\Sigma$-type) sum rule that holds at odd (even) order $b$ automatically holds at order $b+1$.

Using the definitions in Eqs.~\eqref{eq:asdef} and \eqref{eq:deltasigmadef}, we can express $\Delta_i$ and $\Sigma_i$ in terms of the $a$- and $s$-type amplitudes:
\begin{equation}
    \Delta_i = \text{Re}[s_i a_i^*]\,,\qquad
    \Sigma_i = \frac{1}{2}(|s_i|^2 + |a_i|^2)\,.
    \label{eq:as-to-deltasigma}
\end{equation}
Amplitude-squared sum rules can therefore be derived directly from $a$($s$)-type amplitude sum rules. The general idea is to manipulate amplitude sum rules by multiplying them by one another and rewriting the resulting equations in terms of bilinears of $a$- and $s$-type amplitudes. If the resulting equations, which are linear in these bilinears, can be manipulated to eliminate cross terms involving amplitudes corresponding to different channels, one obtains relations involving only bilinears within the same channel. We proceed to describe the particular procedure that we use to realize this idea. We first ``square'' the amplitude sum rules to yield sum rules among $|a_i|^2$ and $|s_i|^2$. We then explain how these relations map onto $\Delta$- and $\Sigma$-type amplitude-squared sum rules.

\paragraph{The ``squaring'' procedure} Next we briefly describe the procedure that allows one to derive sum rules between $|a_i|^2$ from sum rules between $a_i$. Exactly the same construction applies to $|s_i|^2$ and $s_i$. First, we write the full set of $a$-type amplitude sum rules at a given order in matrix form as
\begin{equation}\label{eq:SRs-to-square}
    M\, \vec{a} = 0\,,
\end{equation}
where $\vec{a}$ is the vector of $a$-type amplitudes and $M$ is a matrix of numerical coefficients. We then perform Gaussian elimination to bring the system to reduced row echelon form\footnote{That is, after elementary row operations and a suitable ordering of columns, the matrix is brought, schematically, to the following form
\[
M \;\to\; \begin{pmatrix}
I & -R
\end{pmatrix}\,,
\] where $I$ is the identity matrix.}. This allows us to separate the $a$-type amplitudes into a vector of dependent amplitudes, $\vec{a}_{\rm dep}$, and a vector of the remaining amplitudes, which we refer to as residual amplitudes, $\vec{a}_{\rm res}$, such that
\begin{equation}
    \vec{a}_{\rm dep}=R\,\vec{a}_{\rm res}\,,
\end{equation}
where $R$ is the resulting matrix of coefficients. Squaring these relations, that is, multiplying each equation by its complex conjugate, yields squared dependent amplitudes on the left-hand side and a combination of squared residual amplitudes and their interference terms on the right-hand side. Separating diagonal and interference terms, we obtain
\begin{equation}\label{eq:SR-squared}
    \begin{pmatrix}
    |a_{i_1}|^2\\
    \vdots\\
    |a_{i_{n_\text{dep}}}|^2
    \end{pmatrix}
    =
    R_D
    \begin{pmatrix}
    |a_{j_1}|^2\\
    \vdots\\
    |a_{j_{n_\text{res}}}|^2
    \end{pmatrix}
    +
    R_I
    \begin{pmatrix}
    \vdots\\
    \text{Re}(a_{j}a_{\ell}^\ast)\\
    \vdots
    \end{pmatrix},
\end{equation}
where $a_{i_1},\ldots,a_{i_{n_\text{dep}}}$ and $a_{j_1},\ldots,a_{j_{n_\text{res}}}$ denote the dependent and residual amplitudes, respectively, and the last column vector collects all interference terms between residual amplitudes. The matrices $R_D$ and $R_I$ are obtained from $R$ and give the coefficients of the diagonal and interference terms, respectively. Amplitude-squared sum rules are then obtained by taking linear combinations of the equations in Eq.~\eqref{eq:SR-squared} for which the interference terms cancel. Equivalently, for any vector $\lambda$ in the null space of $R_I^T$, that is,
\begin{equation}
    R_I^T \lambda = 0,
\end{equation}
one obtains
\begin{equation}
    \lambda^T
    \begin{pmatrix}
    |a_{i_1}|^2\\
    \vdots\\
    |a_{i_{n_\text{dep}}}|^2
    \end{pmatrix}
    -
    \lambda^T R_D
    \begin{pmatrix}
    |a_{j_1}|^2\\
    \vdots\\
    |a_{j_{n_\text{res}}}|^2
    \end{pmatrix}
    =0\,,
\end{equation}
which is a sum rule between squared $a$-type amplitudes that holds to the same order in the symmetry breaking as the sum rules in Eq.~\eqref{eq:SRs-to-square}.

\paragraph{Converting from sum rules between $|a|^2$ and $|s|^2$ to $\Delta$ and $\Sigma$}

It is straightforward to see that, under the conditions stated below, any sum rule among squared $a$($s$)-type amplitudes implies a sum rule among the corresponding $\Delta$($\Sigma$)-type amplitudes-squared, with the same numerical coefficients and holding to the same order in the symmetry breaking.

Consider a sum rule between $|a_i|^2$ obtained by applying the ``squaring'' procedure described above to a system of $a$-type amplitude sum rules that hold at order $b$. If the same system of amplitude sum rules, but with all $a_i \to s_i$, holds at order $b-1$, then
\begin{equation}\label{eq:from-a-to-Delta}
    \sum_i \gamma_i |a_i|^2 = \mathcal{O}(\varepsilon^{b+1})
    \qquad \implies \qquad
    \sum_i \gamma_i \Delta_i = \mathcal{O}(\varepsilon^{b+1}) \,.
\end{equation}
This is because if the two systems of amplitude sum rules have the same coefficients, the same linear combinations that eliminate the cross terms when the $a$-type relations are squared also eliminate the cross terms when the corresponding $s$-type relations are multiplied by the complex conjugates of the $a$-type relations. The surviving diagonal terms are $\text{Re}[s_i a_i^*] = \Delta_i$. Since the $a_i = \mathcal{O}(\varepsilon)$, the order-$(b-1)$ accuracy of the $s$-type relations is sufficient for the resulting mixed relation to hold at $\mathcal{O}(\varepsilon^{b+1})$.

Next, consider a sum rule between $|s_i|^2$ that holds at order $b$. If the same sum rule, but with all $|s_i|^2 \to |a_i|^2$, holds at order $b-1$, then
\begin{equation}\label{eq:from-s-to-Sigma}
    \sum_i \gamma_i |s_i|^2 = \mathcal{O}(\varepsilon^{b+1})
    \qquad \implies \qquad
    \sum_i \gamma_i \Sigma_i = \mathcal{O}(\varepsilon^{b+1}) \,.
\end{equation}
For an $s$-type sum rule, $b$ is odd. Since $|a_i|^2$ contains only even powers of $\varepsilon$, the order-$(b-1)$ relation among $|a_i|^2$ also holds at order $b$. Adding it to the relation among $|s_i|^2$ and using Eq.~\eqref{eq:as-to-deltasigma} yields the implication in Eq.~\eqref{eq:from-s-to-Sigma}.

For systems with at least one doublet, the conditions for Eqs.~\eqref{eq:from-a-to-Delta} and \eqref{eq:from-s-to-Sigma} are always satisfied because the lattice construction guarantees a nested structure of amplitude sum rules. Every $b$-dimensional subspace contains a set of $(b-1)$-dimensional subspaces. Adding together the order-$(b-1)$ amplitude sum rules associated with all such subspaces yields an order-$(b-1)$ amplitude sum rule whose coefficients match those of the order-$b$ amplitude sum rule associated with the full $b$-dimensional subspace, but with $a_i \leftrightarrow s_i$. This provides the lower-order $s$-type amplitude sum rules required for Eq.~\eqref{eq:from-a-to-Delta}. Applying the same ``squaring'' procedure to the corresponding lower-order $a$-type amplitude sum rules yields the lower-order sum rules between $|a_i|^2$ required for Eq.~\eqref{eq:from-s-to-Sigma}. However, for systems without a doublet, the lattice construction does not apply directly, and these conditions are not guaranteed to be satisfied. Thus, to use the implications in Eqs.~\eqref{eq:from-a-to-Delta} and \eqref{eq:from-s-to-Sigma} for such systems, it is necessary to check explicitly if the required order-$(b-1)$ $s$-type amplitude sum rules in the first case and the required order-$(b-1)$ sum rules between $|a_i|^2$ in the second case exist.

\texttt{FlaSR} implements the ``squaring'' procedure described above to derive sum rules between $|a_i|^2$ and $|s_i|^2$. It checks to ensure that the conditions for Eqs.~\eqref{eq:from-a-to-Delta} and \eqref{eq:from-s-to-Sigma} hold before applying them to yield $\Delta$($\Sigma$)-type amplitude-squared sum rules. These relations can then be readily converted into sum rules between squared $A$-type amplitudes, using the definitions in Eq.~\eqref{eq:deltasigmadef}, or into rate sum rules.

\subsection{Scope of applications and results}
\label{sec:scope}

We conclude this section by discussing the types of systems to which \texttt{FlaSR} can be applied and the scope of its results.

\texttt{FlaSR}'s algorithm relies on the following assumptions:
\begin{itemize}
    \item \textit{Pure $SU(2)$ irreps.} We assume that all external particles and the operators in the symmetry-limit Hamiltonian are given by components of irreducible representations of $SU(2)$. In particular, the Hamiltonian is assumed to be dominated by operators with one fixed value of total $U$-spin; see Eq.~\eqref{eq:Heff0}.

    Consequently, the present version of the package does not cover cases in which external particles are mixtures of irreps, nor cases in which several values of total $U$-spin must be included in the Hamiltonian. An example of the first type is provided by physical neutral mesons such as $\pi^0$, $\eta$, and $\eta'$, which are mixtures rather than pure $U$-spin states. An example of the second type is given by hadronic charm decays when CP violation is taken into account. In this case the full Hamiltonian contains both triplet and singlet contributions. Such situations require the direct-sum formalism developed in Ref.~\cite{gavrilova2024direct}, which is not currently implemented in \texttt{FlaSR}.

    \item \textit{Completeness of the system.} We also assume that the system is complete, i.e.\ that every channel allowed by pure $SU(2)$ group theory is physically realized. Physically, this requires that all corresponding processes have sufficient phase space and that none of the coefficients $f_{u,m}$ in Eq.~\eqref{eq:Heff0} vanish. The latter assumption can fail, for example, in isospin- or $V$-spin-related systems of hadronic weak decays, where some channels allowed by the flavor symmetry are forbidden by electric-charge conservation.
\end{itemize}

Next, we comment on cases in which \texttt{FlaSR} can be applied beyond weak decays related by $U$-spin:
\begin{itemize}
    \item \textit{Semileptonic weak decays related by isospin or $V$-spin.} In semileptonic systems, the leptons, which are flavor singlets, can account for the electric charge, so all coefficients $f_{u,m}$ in Eq.~\eqref{eq:Heff0} are nonzero and the system is complete, provided all channels also have sufficient phase space. An explicit example is the semileptonic $K\to\pi$ system discussed in Section~V.C.2 of Ref.~\cite{gavrilova2022structure}.
    
    \item \textit{Scattering and decays mediated by the strong interaction for isospin, $U$-spin, and $V$-spin.} In this case, the symmetry-limit Hamiltonian is a singlet. The package can therefore be applied whenever the full symmetry-related set of channels is kinematically allowed.
    
    \item \textit{Mass sum rules.} Hadron masses can be treated within the same group-theoretical framework as matrix elements between one-hadron states, i.e.\ as $h \to h$ amplitudes. The package can therefore also be used to derive sum rules among masses of hadrons that belong to isospin, $U$-spin, or $V$-spin irreps. We provide an example in Section~\ref{sec:mass sum rules}.
\end{itemize}

Finally, we briefly comment on the scope of \texttt{FlaSR}'s output. For a given system, the program generates a complete set of linearly independent amplitude sum rules through all orders of the symmetry breaking. However, the current ``squaring'' procedure is not guaranteed to generate a complete set of linearly independent amplitude-squared sum rules for the system. A systematic characterization of the full space of amplitude-squared sum rules requires future theoretical work.

\section{Instructions for using the package}
\label{sec:instructions}

This section describes the basic workflow for using \texttt{FlaSR}. At the user level, the package is organized around two functions. The function \texttt{generateSRs} constructs the system and computes the amplitude and amplitude-squared sum rules. This function's output is formatted by the \texttt{printSystem} function for display. Throughout this section, we use the abstract group-theoretic system that matches the structure of $D^0 \to P^+ P^-$, defined in Eq.~\eqref{eq:DPPmultiplet} with a triplet Hamiltonian, as a running example.

\subsection{Installation}

\texttt{FlaSR} was developed and tested on Mathematica~14.3 and Python~3.13. The current version can be downloaded at
\begin{center}
    \url{https://github.com/Flavor-Sum-Rules/FlaSR}.
\end{center}

To load the package, place \verb|FlaSR.m| and \verb|FlaSR.py| in the same directory as the Mathematica notebook. In the notebook, set the working directory to the notebook's directory and load the Mathematica package:
\begin{Verbatim}[frame=single,commandchars=\\\{\},codes={\catcode`$=3\catcode`^=7\catcode`_=8}]
SetDirectory[NotebookDirectory[]];
Get["FlaSR.m"];
\end{Verbatim}

The Mathematica package uses Python code from the \verb|FlaSR.py| file to perform part of the calculations. Start an external Python session and load the \verb|FlaSR.py| file into the session:
\begin{Verbatim}[frame=single,commandchars=\\\{\},codes={\catcode`$=3\catcode`^=7\catcode`_=8}]
session = StartExternalSession["Python"];
startPythonSession[session, "FlaSR.py"];
\end{Verbatim}

If the Mathematica kernel is restarted, these initialization commands must be rerun.

\subsection{Generating the system and sum rules}

\begin{table}[]
    \centering
    \begin{tabular}{|p{0.18\linewidth}|p{0.17\linewidth}|p{0.55\linewidth}|}
        \hline
        \textbf{Variable} & \textbf{Type (Default)} & \textbf{Description} \\ \hline\hline
        
        \multicolumn{3}{|l|}{\textit{Arguments}} \\ \hline
        \verb|in| & List & In state irreps in abstract mode, or particle multiplets in physical mode. \\ \hline
        \verb|h| & List & Hamiltonian irrep in abstract mode, or ordered list of Hamiltonian coefficients in physical mode. \\ \hline
        \verb|out| & List & Out state irreps in abstract mode, or particle multiplets in physical mode. \\ \hline\hline
    
        \multicolumn{3}{|l|}{\textit{Options}} \\ \hline
        \verb|phys| & Boolean (\verb|False|) & Switches between abstract irrep input (\verb|False|) and physical multiplet input (\verb|True|) modes. \\ \hline
        \verb|obs| & String (\verb|"Diff"|) & Indicates whether observables are differential (\verb|"Diff"|) or integrated (\verb|"Int"|). Only applicable when \verb|phys -> True|. \\ \hline\hline
    
        \multicolumn{3}{|l|}{\textit{Return values}} \\ \hline
        \verb|system| & Association & Stores the system's irreps, amplitudes, amplitude sum rule matrices, and amplitude-squared sum rule matrices. \\ \hline
    \end{tabular}
    \caption{Interface of \texttt{generateSRs}. Usage details of the \texttt{obs} option are discussed in Section~\ref{sec:examples}. The keys and values of the \texttt{system} association are detailed in Table~\ref{tab:generateSRs output}.}
    \label{tab:generateSRs}
\end{table}

Mathematically, every \U-spin system is defined by specifying the irreps in the external states and the Hamiltonian. To generate all amplitudes, amplitude sum rules, and amplitude-squared sum rules for a \U-spin system, use
\begin{center}
    \texttt{generateSRs[\{in reps\}, \{H rep\}, \{out reps\}, \ldots]}\,.
\end{center}
This function requires the three lists \verb|{in reps}|, \verb|{H rep}|, \verb|{out reps}| as inputs, with the dots indicating additional input options which can be set via \verb|option -> value|. We summarize the \verb|generateSRs| interface in Table~\ref{tab:generateSRs}. There are two input modes for this function:
\begin{enumerate}
    \item physical (\verb|phys -> True|), and
    \item abstract (\verb|phys -> False|).
\end{enumerate}
The physical input mode is discussed in detail in Section~\ref{sec:examples}. Here we focus on the abstract mode, which is the default input mode of \verb|generateSRs|. In this mode, the inputs are lists of irreps in the initial state, the Hamiltonian, and the final state, where each irrep is labeled by its total $U$-spin angular momentum. For the running example, $D^0\to P^+ P^-$, the system is defined by a \U-spin singlet ($u_{D^0} = 0$) in the initial state, a triplet ($u_H = 1$) in the Hamiltonian, and two doublets ($u_{P^+} = 1/2$, $u_{P^-} = 1/2$) in the final state. The amplitudes and sum rules for this system are generated as
\begin{Verbatim}[frame=single,commandchars=\\\{\},codes={\catcode`$=3\catcode`^=7\catcode`_=8}]
system = generateSRs[\{0\}, \{1\}, \{1/2, 1/2\}];
\end{Verbatim}
We note that in the abstract input mode  \texttt{FlaSR} treats all irreps as distinguishable, even when some of them carry the same value of $U$-spin. In particular, in the example above the two doublets in the final state are treated as two distinct multiplets. By contrast, in the physical input mode \texttt{FlaSR} can account for the presence of identical multiplets in the final state. We discuss this feature in detail in Section~\ref{sec:examples}.

The output of the \verb|generateSRs| function is a Mathematica association containing the system's representations, amplitudes, and sum rules. A complete list of keys and values for the association outputted by this function is provided in Table~\ref{tab:generateSRs output}. The most commonly used association keys are \verb|"Amplitudes"|, \verb|"n amps"|, \verb|"ASRs"|, and \verb|"A2SRs"|. The first two of these store the amplitude table and number of amplitudes, while \texttt{"ASRs"} and \texttt{"A2SRs"} store amplitude and amplitude-squared sum rules respectively. As we explain below, sum rules are stored in the form of matrices of coefficients, which conveniently enables sum rules to be extracted and manipulated.

\begin{table}[]
    \centering
    \begin{tabular}{|p{0.18\linewidth}|p{0.17\linewidth}|p{0.55\linewidth}|}
        \hline
        \textbf{Key} & \textbf{Type (Default)} & \textbf{Value} \\ \hline
        \verb|"Amplitudes"| & List & All amplitudes in the system. Each element of the \texttt{"Amplitudes"} list is an association corresponding to an amplitude pair. Each association has the following keys and values:
        \vspace{-0.75em}\begin{itemize}[leftmargin=*,itemsep=-0.25em]
            \item \verb|"Process"| (List): Physical processes for the amplitude pair. Only appears for physical systems.
            \item \verb|"QN label"| (List): $m$-QN labels for the amplitude pair.
            \item \verb|"n-tuple"| (List): $n$-tuples for the amplitude pair.
            \item \verb|"Coord"| (String): Coordinate for the amplitude pair.
            \item \verb|"Binary index"| (List): Binary indices for the amplitude pair.
            \item \verb|"mu"| (Expression): $\mu$-factor for the coordinate.
            \item \verb|"CG"| (Expression): Clebsch--Gordan symmetrization coefficient.
            \item \verb|"CKM"| (List): CKM factors in the Hamiltonian. Only appears for physical systems.
            \item \verb|"Integrated channel ID"| (List): Unique integrated channel numbers for the amplitude pair. Only appears for physical systems with integrated observables.
        \end{itemize} \\ \hline
        \verb|"n amps"| & Integer & Number of amplitudes in the system. \\ \hline
        \verb|"ASRs"| & List & Amplitude sum rule matrices in the $a$($s$)-type amplitude bases. \\ \hline
        \verb|"A2SRs"| & List & Amplitude-squared sum rule matrices in the $\Delta$($\Sigma$)-type amplitude-squared bases. \\ \hline
        \verb|"n ASRs"| & List & Amplitude sum rule counts at each $b$. \\ \hline
        \verb|"n A2SRs"| & List & Amplitude-squared sum rule counts at each $b$. \\ \hline
        \verb|"Irreps"| & List & Irreps in the system. \\ \hline
        \verb|"Multiplets"| & List & Multiplets in the physical system. Empty when using the abstract input mode. \\ \hline
        \verb|"n doublets"| & Integer & Number of would-be doublets, $n$, for the system. \\ \hline
        \verb|"p factor"| & Integer & $(-1)^p$ factor for defining $a$($s$)-type amplitudes. \\ \hline
        \verb|"SR extract"| & List (\verb|None|) & Sum rule matrices extracted according to user-selected options of \verb|printSystem|. \\ \hline
        \verb|"Amp vector"| & List (\verb|None|) & Vector of formatted $A$-type amplitudes (or $|A|^2$ amplitudes-squared) or vectors of formatted $a$($s$)-type amplitudes (or $\Delta$($\Sigma$)-type amplitudes-squared). \\ \hline
    \end{tabular}
    \caption{Summary of the \texttt{system} association outputted by \texttt{generateSRs}; see \href{https://github.com/Flavor-Sum-Rules/FlaSR}{GitHub} for more details on the structure of the contents. The technical details of binary indices, $\mu$-factors, and Clebsch--Gordan symmetrization coefficients are given in Ref.~\cite{gavrilova2022structure}. These fields are maintained for internal use and are hidden from the default output of the \texttt{printSystem} function in Table~\ref{tab:printSystem}. Integrated channel numbers are discussed in Section~\ref{sec:examples}. The \texttt{"SR extract"} and \texttt{"Amp vector"} keys, which allow the user to extract particular sum rules for manipulation, are initialized to the value \texttt{None} by \texttt{generateSRs} and are redefined after running \texttt{printSystem}.}
    \label{tab:generateSRs output}
\end{table}

\begin{table}[]
    \centering
    \begin{tabular}{|p{0.13\linewidth}|p{0.2\linewidth}|p{0.57\linewidth}|}
        \hline
        \textbf{Variable} & \textbf{Type (Default)} & \textbf{Description} \\ \hline\hline
        
        \multicolumn{3}{|l|}{\textit{Arguments}} \\ \hline
        \verb|system| & Association & Output of \verb|generateSRs|. See Table~\ref{tab:generateSRs output} for details. \\ \hline\hline

        \multicolumn{3}{|l|}{\textit{Required options}$^\dag$} \\ \hline
        \verb|ampType| & List (\verb|None|) & Amplitude basis: use \verb|{A}| for $A$-type amplitudes or \verb|{a,s}| for $a$($s$)-type amplitudes. \\ \hline
        \verb|amp2Type| & List (\verb|None|) & Amplitude-squared basis: use \verb|{A}| for $|A|^2$ amplitudes-squared, \texttt{\{$\Delta$,$\Sigma$\}} for $\Delta$($\Sigma$)-type amplitudes-squared, or \texttt{\{$\Gamma$\}} for rates. When using \texttt{\{$\Gamma$\}}, make sure to also set \texttt{amp2Quad -> True}. \\ \hline\hline

        \multicolumn{3}{|l|}{\textit{Options}} \\ \hline
        \verb|ampFormat| & String (\verb|"n-tuple"|, except for physical systems with $A$-type amplitudes, then \verb|"Process"|) & Amplitude labeling convention: physical processes (\texttt{"Process"}, only available for $A$-type amplitudes), QNs (\texttt{"QN label"}), $n$-tuples (\texttt{"n-tuple"}), coordinates (\texttt{"Coord"}, only available for $a$($s$)-type amplitudes), binary indices (\texttt{"Binary index"}), or user-defined labels. \\ \hline
        \verb|expandSRs| & Boolean (\verb|False|) & Indicates whether to display sum rules as expanded algebraic expressions of amplitudes (\verb|True|) or to keep them as coefficient matrices (\verb|False|). \\ \hline
        \verb|CKM| & Boolean (\verb|False|) & Indicates whether to display CKM factors in sum rules. Amplitudes and rates should be interpreted as physical quantities when \verb|CKM -> True| and CKM-normalized quantities when \verb|CKM -> False|. \\ \hline
        \verb|amp2Quad| & Boolean (\verb|False|) & Indicates whether the symbol assigned to \verb|amp2Type| is quadratically (\verb|True|) or linearly (\verb|False|) dependent on $A$. When \texttt{amp2Quad -> True}, use a symbol representing a quantity that is proportional to $|A|^2$, e.g., \texttt{amp2Type -> \{$\Gamma$\}} for rates. \\ \hline
        \verb|showReps| & Boolean (\verb|True|) & Indicates whether to print the system summary.\\ \hline
        \verb|showAmps| & Boolean (\verb|True|) & Indicates whether to print the amplitude table. \\ \hline
        \verb|showFactors| & Boolean (\verb|False|) & Indicates whether to print binary indices, $\mu$, and CG in the amplitude table. \\ \hline
        \verb|showASRs| & Boolean (\verb|True|) & Indicates whether to print amplitude sum rules. \\ \hline
        \verb|showA2SRs| & Boolean (\verb|True|) & Indicates whether to print amplitude-squared sum rules. \\ \hline
        \verb|b| & All, Integer, or Range specification (\verb|All|) & Breaking order(s) at which to print sum rules: all possible orders (\verb|All|), at a particular order $b_0$ (\texttt{$\texttt{b}_0$}), or over a range (\verb|{min,max,increment=1}|). \\ \hline\hline

        \multicolumn{3}{|l|}{\textit{Return values}} \\ \hline
        \verb|system| & Association & The inputted \verb|system| association, modified to include the last printed set of sum rule matrices and vector(s) of amplitudes (or amplitudes-squared). For example, if \verb|printSystem| is used to print both amplitude and amplitude-squared sum rules, then \verb|"SR extract"| and \verb|"Amp vector"| will respectively contain amplitude-squared sum rules and amplitudes-squared. Modified keys and values: \vspace{-0.75em}\begin{itemize}[leftmargin=*,itemsep=-0.25em]
            \item \verb|"SR extract"| (List): Sum rule coefficient matrices.
            \item \verb|"Amp vector"| (List): Vector(s) of formatted amplitudes (or amplitudes-squared).
        \end{itemize} \\ \hline
    \end{tabular}
    \caption{Interface of \texttt{printSystem}. $^\dag$These options are required unless the corresponding sum rules are explicitly suppressed from display via \texttt{showASRs -> False} or \texttt{showA2SRs -> False}.}
    \label{tab:printSystem}
\end{table}

To access a particular key of the \texttt{system} association, use the syntax \verb|system[["Key"]]|. For example, 
\begin{Verbatim}[frame=single,commandchars=\\\{\},codes={\catcode`$=3\catcode`^=7\catcode`_=8}]
system[["Amplitudes"]]
\end{Verbatim}
allows access to the \verb|"Amplitudes"| key, which contains information about all amplitudes of the system; see Table~\ref{tab:generateSRs output} for details.

\subsection{Formatting and interpreting the output}

The function
\begin{center}
    \texttt{printSystem[system, ampType -> \{a,s\} or \{A\},}
    \texttt{amp2Type -> \{$\Delta$,$\Sigma$\} or \{A\}, \ldots]}
\end{center}
formats the contents of \verb|system| for display. The interface for \verb|printSystem|, including the full list of options, is summarized in Table~\ref{tab:printSystem}. This function takes in the \verb|system| association outputted by \verb|generateSRs|, along with the options \verb|ampType| and \verb|amp2Type| for selecting amplitude and amplitude-squared bases. The \verb|ampType| option takes in $1$- or $2$-element lists of symbol(s) to be used to denote $a$($s$)-type or $A$-type amplitudes in the displayed output. Although we have indicated using the symbols \verb|a|, \verb|s|, and \verb|A|, these are simply suggestions to match the conventions of Section~\ref{sec:theory}; other symbols can be used in their place with the understanding that the first (second) symbol in a $2$-element list will be used by \verb|FlaSR| to denote an $a$-type ($s$-type) amplitude, and the symbol in a $1$-element list will be used to denote an $A$-type amplitude. For \verb|amp2Type|, the same applies to the suggestions of the \texttt{$\Delta$}, \texttt{$\Sigma$}, and \texttt{A} symbols corresponding to the quantities $\Delta$, $\Sigma$, and $|A|^2$ respectively; see Table~\ref{tab:printSystem} and Section~\ref{sec:B to DOO} for details on how to write amplitudes-squared in terms of quantities quadratically dependent on $A$, such as $\Gamma$ for rates. By default, both \verb|ampType| and \verb|amp2Type| are required input options unless the user explicitly suppresses the corresponding sum rules from display; see Table~\ref{tab:printSystem} for details.

There are a number of supplemental options for formatting the output as well. Of these, the most frequently used options are \verb|ampFormat| for selecting the amplitude labeling convention, \verb|expandSRs| for choosing whether to display the sum rules as expanded algebraic expressions or as matrices, and \verb|CKM| for indicating whether to restore CKM factors. Amplitudes, amplitudes-squared, and rates should be interpreted as physical quantities---i.e. $\mathcal{A}_i$, $|\mathcal{A}_i|^2$, and $\Gamma_i$---when \verb|CKM -> True| and CKM-normalized quantities---i.e. $A_i \equiv \mathcal{A}_i/f_{u,m}$, $|A_i|^2 \equiv |\mathcal{A}_i/f_{u,m}|^2$, and $\Gamma_i/|f_{u,m}|^2$---when \verb|CKM -> False|.

Here, we choose to display amplitude sum rules using $a$($s$)-type amplitudes (\texttt{ampType -> \{a,s\}}), defined in Eq.~\eqref{eq:asdef}, and amplitude-squared sum rules using $\Delta$($\Sigma$)-type amplitudes-squared (\texttt{amp2Type -> \{$\Delta$,$\Sigma$\}}\footnote{To enter the $\Delta$ symbol in Mathematica, press \texttt{Esc} before and after entering \texttt{Delta} or type $\texttt{\textbackslash[CapitalDelta]}$, and similarly for the $\Sigma$ symbol.}), defined in Eq.~\eqref{eq:deltasigmadef}:
\begin{Verbatim}[frame=single,commandchars=\\\{\},codes={\catcode`$=3\catcode`^=7\catcode`_=8}]
printSystem[system, ampType -> \{a,s\}, amp2Type -> \{$\Delta$,$\Sigma$\}];
\end{Verbatim}

The resulting program output has three parts that we discuss below: system summary, amplitude table, and sum rules.
\begin{myframe}
$\left.
\begin{minipage}{0.6\textwidth}
\begin{Verbatim}[vspace=0em,xleftmargin=-2.75mm,commandchars=\\\{\},codes={\catcode`$=3\catcode`^=7\catcode`_=8\catcode`&=4}]
System: \{0,$\frac{1}{2}$,$\frac{1}{2}$,1\}
-------------------------
Number of would-be doublets: 4
In: \{0\}
H: \{1\}
Out: \{$\frac{1}{2}$,$\frac{1}{2}$\}
\end{Verbatim}
\end{minipage}
\qquad\qquad\right\} 1$
\begin{Verbatim}[vspace=0.5em,xleftmargin=-2.75mm]
-------------------------
\end{Verbatim}
$\left.
\begin{minipage}{0.6\textwidth}
\begin{Verbatim}[vspace=-0.25em,xleftmargin=-2.75mm,commandchars=\\\{\},codes={\catcode`$=3\catcode`^=7\catcode`_=8\catcode`&=4}]
Amplitude table
Number of amplitudes: 4
a/s definitions: $\texttt{a}_\texttt{i} = \texttt{A}_\texttt{i} - \overline{\texttt{A}}_\texttt{i}$, $\texttt{s}_\texttt{i} = \texttt{A}_\texttt{i} + \overline{\texttt{A}}_\texttt{i}$
$\Delta$/$\Sigma$ definitions: $\Delta_\texttt{i} = |\texttt{A}_\texttt{i}|^2 - |\overline{\texttt{A}}_\texttt{i}|^2$, $\Sigma_\texttt{i} = |\texttt{A}_\texttt{i}|^2 + |\overline{\texttt{A}}_\texttt{i}|^2$
$\begin{matrix*}[l]\texttt{QN label} && \texttt{n-tuple} && \texttt{Coord}\\0 \overset{-1}{\to} -\frac{1}{2} -\frac{1}{2} && \texttt{(-,-,++)} && \texttt{(1)}\\0 \overset{+1}{\to} +\frac{1}{2} +\frac{1}{2} && \texttt{(+,+,--)} && \\\\0 \overset{0}{\to} -\frac{1}{2} +\frac{1}{2} && \texttt{(-,+,-+)} && \texttt{(2)}\\0 \overset{0}{\to} +\frac{1}{2} -\frac{1}{2} && \texttt{(+,-,-+)} && \end{matrix*}$
\end{Verbatim}
\end{minipage}
\qquad\qquad\right\} 2$
\begin{Verbatim}[vspace=0.5em,xleftmargin=-2.75mm]
-------------------------
\end{Verbatim}
$\left.
\begin{minipage}{0.6\textwidth}
\begin{Verbatim}[vspace=-0.25em,xleftmargin=-2.75mm,commandchars=\\\{\},codes={\catcode`$=3\catcode`^=7\catcode`_=8\catcode`&=4}]
Amplitude sum rules
b = 0
Number of SRs: 2
$\begin{pmatrix}\texttt{a}_{(-,-,++)} & \texttt{a}_{(-,+,-+)}\\\hline 1 & 0\\0 & 1\end{pmatrix}$
b = 1
Number of SRs: 1
$\begin{pmatrix}\texttt{s}_{(-,-,++)} & \texttt{s}_{(-,+,-+)}\\\hline 1 & -\sqrt{2}\end{pmatrix}$
-------------------------
Amplitude-squared sum rules
b = 0
Number of SRs: 2
$\begin{pmatrix}\Delta_{(-,-,++)} & \Delta_{(-,+,-+)}\\\hline 1 & 0\\0 & 1\end{pmatrix}$
b = 1
Number of SRs: 1
$\begin{pmatrix}\Sigma_{(-,-,++)} & \Sigma_{(-,+,-+)}\\\hline 1 & -2\end{pmatrix}$
\end{Verbatim}
\end{minipage}
\qquad\qquad\right\} 3$
\end{myframe}

We comment on each part of the output:
\begin{enumerate}
    \item The \textit{system summary} lists the irreps (\texttt{System}) and their assignments to the initial state~(\texttt{In}), Hamiltonian~(\texttt{H}), and the final state~(\texttt{Out}) in the order used internally by the package. Irreps are ordered by increasing total \U-spin, and for physical systems, ties are broken by their order of appearance in the user input. The total number of would-be doublets $n$ is also shown.
    \item The \textit{amplitude table} lists the total number of amplitudes in the system, definitions for $a$($s$)-type amplitudes and $\Delta$($\Sigma$)-type amplitudes-squared according to the definitions in Eq.~\eqref{eq:asdef} and Eq.~\eqref{eq:deltasigmadef}, and a table of the amplitudes written in QN (\texttt{QN label}), $n$-tuple (\texttt{n-tuple}), and coordinate (\texttt{Coord}) notations. The $n$-tuple and coordinate notations are defined using the internal irrep order, while the QN notation maintains the order of the user input. For physical systems, where \texttt{phys -> True}, there are two additional columns listing the physical processes (\texttt{Process}) and CKM factors (\texttt{CKM}) corresponding to each amplitude; see Section~\ref{sec:examples} for examples. Each row of the table corresponds to a unique amplitude; amplitude pairs are grouped together. Thus, a single coordinate is used to refer to both an amplitude and its conjugate in the row below.
    \item The \textit{sum rules} section presents the amplitude and amplitude-squared sum rules for the system. Each order of breaking $b$ is accompanied by a count of the number of sum rules at that order along with a matrix of sum rule coefficients. The sum rules listed at a given order $b$ should be interpreted as the relations that are broken by corrections of $\mathcal{O}(\varepsilon^{b+1})$. The header row of a sum rule coefficient matrix lists the amplitudes in the system, and the remaining rows correspond to sum rules. Multiplying the amplitudes by their respective coefficients along a row and setting the resulting expression equal to 0 constitutes a sum rule. By default, the program indexes amplitudes in abstract group-theoretic systems by their $n$-tuples and amplitudes in physical systems by their physical processes.
\end{enumerate}

We now explicitly interpret the output for this example. There are two amplitude sum rules valid up to $b = 0$ that are broken by corrections of $\mathcal{O}(\ve)$. Writing these using both $a$($s$)-type and $A$-type amplitudes, the amplitude sum rules at $b = 0$ are
\begin{align}
    a_{(-,-,++)} &= A_{(-,-,++)} - A_{(+,+,--)} = \mathcal{O}(\ve)\,, \nonumber\\
    a_{(-,+,-+)} &= A_{(-,+,-+)} - A_{(+,-,-+)} = \mathcal{O}(\ve) \label{eq:instructions asrb0}
\,.\end{align}
There is a single amplitude sum rule holding at $b = 1$:
\begin{align}
    s_{(-,-,++)} - \sqrt{2}s_{(-,+,-+)} = A_{(-,-,++)} + A_{(+,+,--)} - \sqrt{2}(A_{(-,+,-+)} + A_{(+,-,-+)}) = \mathcal{O}(\ve^2)
\,.\end{align}

The output for sum rules between squared amplitudes is interpreted in the exact same way. There are two amplitude-squared sum rules holding up to $b = 0$:
\begin{align}
    \Delta_{(-,-,++)} &= |A_{(-,-,++)}|^2 - |A_{(+,+,--)}|^2 = \mathcal{O}(\ve)\,, \nonumber\\
    \Delta_{(-,+,-+)} &= |A_{(-,+,-+)}|^2 - |A_{(+,-,-+)}|^2 = \mathcal{O}(\ve)
\,.\end{align}
Finally, there is one amplitude-squared sum rule at $b = 1$:
\begin{align}
    \Sigma_{(-,-,++)} - 2\Sigma_{(-,+,-+)} = |A_{(-,-,++)}|^2 + |A_{(+,+,--)}|^2 - 2(|A_{(-,+,-+)}|^2 + |A_{(+,-,-+)}|^2) = \mathcal{O}(\ve^2) \label{eq:instructions a2srb1}
\,.\end{align}
While this example's results are group-theoretic, they hold for any \U-spin system with a matching \U-spin structure. In Section~\ref{sec:D to PP}, we reframe these results in terms of physical amplitudes when we discuss the physical $D^0 \to P^+ P^-$ system.\\

We conclude by summarizing the basic \texttt{FlaSR} workflow:
\begin{enumerate}
    \item Load the Mathematica package and initialize the external Python session.
    \item Call \texttt{generateSRs} to construct the system and compute its sum rules. The program will return a Mathematica association containing information about the system's representations, amplitudes, and sum rules.
    \item Call \texttt{printSystem} to display the output in the preferred amplitude and amplitude-squared bases and labeling notation. When interpreting results, remember that a sum rule displayed at order $b$ is broken by corrections of $\mathcal{O}(\ve^{b+1})$.
\end{enumerate}

A summary of all functions in the \verb|FlaSR| package is given in Table~\ref{tab:all fns}.

\begin{table}[]
    \centering
    \begin{tabular}{|p{0.20\linewidth}|p{0.73\linewidth}|}
        \hline
        \textbf{Function} & \textbf{Description} \\ \hline\hline

        \multicolumn{2}{|l|}{\textit{Main functions}} \\ \hline
        \texttt{generateSRs} & \verb|generateSRs[in,h,out]| finds amplitudes, amplitude sum rules, and amplitude-squared sum rules for a given system. \\ \hline
        \texttt{printSystem} & \texttt{printSystem[system,ampType -> \{a,s\}/\{A\}, amp2Type -> \{$\Delta$,$\Sigma$\}/\{A\}]} prints information about the system's representations, amplitudes, and sum rules and adds formatted sum rules and amplitude vectors to the system association. \\ \hline\hline

        \multicolumn{2}{|l|}{\textit{Additional functions}} \\ \hline
        \texttt{FlaSRHelp} & \verb|FlaSRHelp[FlaSR function name]| prints extended documentation on a \verb|FlaSR| function's arguments, options, and outputs. \\ \hline
        \texttt{startPythonSession} & \verb|startPythonSession[session,path]| checks for a valid Python session and file path and loads the \verb|FlaSR.py| Python file. \\ \hline
        \texttt{printAmps} & \verb|printAmps[system]| prints the system's amplitudes, $a$($s$)-type amplitude definitions, and $\Delta$($\Sigma$)-type amplitude-squared definitions. \\ \hline
        \texttt{printSRs} & \texttt{printSRs[system,ampType -> \{a,s\}/\{A\} OR amp2Type -> \{$\Delta$,$\Sigma$\}/\{A\}]} prints amplitude (or amplitude-squared) sum rules at each order of breaking and extracts formatted sum rule matrices and amplitude vector(s) for manipulation. \\ \hline
        \texttt{labelAmps} & \verb|labelAmps[system,colName,labels]| adds a column of custom labels to \verb|system[["Amplitudes"]]|. \\ \hline
        \texttt{unlabelAmps} & \verb|unlabelAmps[system,colNames]| removes the selected column(s) from \verb|system[["Amplitudes"]]|. \\ \hline
        \texttt{generateASRs} & \verb|generateASRs[in,h,out]| finds amplitudes and amplitude sum rules for a given system. \\ \hline
        \texttt{findA2SRMat} & \verb|findA2SRMat[ASRMat]| finds the amplitude-squared sum rule matrix for a given amplitude sum rule matrix. \\ \hline
        \texttt{numAmps} & \verb|numAmps[system,nPairs:False]| returns the total number of amplitudes (or amplitude pairs) in the system. \\ \hline
        \texttt{numSRs} & \verb|numSRs[system,squared:False]| returns the number of amplitude (or amplitude-squared) sum rules found at each order of breaking. \\ \hline
    \end{tabular}
    \caption{Summary of functions in the \texttt{FlaSR} package. For extended documentation on syntax and outputs, please refer to the \href{https://github.com/Flavor-Sum-Rules/FlaSR}{GitHub repository}.}
    \label{tab:all fns}
\end{table}

\section{Physical systems}
\label{sec:examples}

For physical systems, it is possible to use \texttt{FlaSR} to write sum rules in terms of physical amplitudes. This is achieved through the \texttt{generateSRs} function's physical input mode, which is enabled by setting the option \texttt{phys -> True}. In this mode, each irrep in the external states is replaced by its corresponding particle multiplet, where each multiplet is entered as a list of particle names ordered from highest to lowest $m$-QN. For example, the $P^+$ doublet defined in Eq.~\eqref{eq:DPPmultiplet} is entered as
\begin{Verbatim}[frame=single,commandchars=\\\{\},codes={\catcode`$=3\catcode`^=7\catcode`_=8}]
Pp = \{"$\texttt{K}^+$","$\pi^+$"\};
\end{Verbatim}
Likewise, the Hamiltonian is entered as the ordered list of CKM factors or other coefficients multiplying the components of the symmetry-limit Hamiltonian. For example, a generic triplet Hamiltonian is entered as
\begin{Verbatim}[frame=single,commandchars=\\\{\},codes={\catcode`$=3\catcode`^=7\catcode`_=8}]
H = \{$\texttt{f}_{1,1}$, $\texttt{f}_{1,0}$, $\texttt{f}_{1,-1}$\};
\end{Verbatim}
where the entries correspond to the $m = +1, 0, -1$ components, respectively; see also Eq.~\eqref{eq:Heff0}. For a singlet Hamiltonian in the physical mode, the corresponding list is simply \verb|H = {1}|; see Section~\ref{sec:mass sum rules} for an example.

\paragraph{Identical multiplets} Before presenting the use cases, we note that for physical systems (\texttt{phys -> True}) with identical multiplets in the final state, one must distinguish between differential and integrated observables in order to interpret the program output correctly. In \texttt{FlaSR}, this distinction is controlled in the physical mode of the \texttt{generateSRs} function through the option \texttt{obs -> "Diff"} or \texttt{"Int"}, where \texttt{obs -> "Diff"} is the default setting. Below, we briefly illustrate both the physical distinction between differential and integrated observables and how it is handled in \texttt{FlaSR} through the \texttt{obs} option. The implementation in \texttt{FlaSR} follows the general discussion of systems with identical multiplets in the final state given in Sec.~2.3 of Ref.~\cite{Gavrilova:2026ryc}; see also Appendix~B therein for an explicit example.

Suppose that, among other particles, the final state of the system contains two copies of the doublet $P^+$, defined in Eq.~\eqref{eq:DPPmultiplet} and denoted by \texttt{Pp} in the code block above. At the level of differential observables, such as differential decay rates and differential cross sections, these two copies are implicitly distinguished by their momenta, so one should think of the final state as containing momentum-labeled multiplets $P^+_{p_1}P^+_{p_2}$. Keeping the rest of the final state and its momentum assignment fixed, the configurations $\pi^+_{p_1}K^+_{p_2}$ and $K^+_{p_1}\pi^+_{p_2}$ are distinct and must therefore appear separately in differential sum rules. In \texttt{FlaSR}, however, momentum labels are always suppressed. Thus, in the \texttt{obs -> "Diff"} mode, these two configurations appear simply as $\pi^+ K^+$ and $K^+ \pi^+$, and the two orderings must be understood as corresponding to different momentum assignments and therefore to different amplitudes, squared amplitudes, and, more generally, differential observables.

For observables integrated over the final-state phase space, such as partial decay widths and integrated cross sections, the momentum labels are no longer part of the observable, and configurations that differ only by permuting the two $P^+$ copies must be identified. Accordingly, the two differential contributions $\pi^+_{p_1}K^+_{p_2}$ and $K^+_{p_1}\pi^+_{p_2}$ correspond to a single integrated $\pi^+K^+$ channel. In addition, channels with two identical particles, such as $\pi^+\pi^+$ and $K^+K^+$, carry the usual combinatorial factors from the standard definition of the phase-space integral for identical particles. In \texttt{FlaSR}, these identifications and combinatorial factors are taken into account automatically when \texttt{obs -> "Int"} is used. For a concrete illustration, see Sec.~\ref{sec:DPPP}, where we also compare the output in the two settings. Finally, we note for completeness that the \texttt{obs} option only affects amplitude-squared sum rules and does not affect amplitude sum rules.\\

We now turn to physical examples illustrating the use of \texttt{FlaSR}. In Section~\ref{sec:D to PP} we use the $D^0\to P^+P^-$ system to demonstrate the basic physical mode workflow, from defining a physical system to interpreting the resulting sum rules. In Section~\ref{sec:DPPP} we consider a three-body charm-decay system with two identical multiplets in the final state and highlight the distinction between \texttt{obs -> "Diff"} and \texttt{obs -> "Int"}. In Section~\ref{sec:B to DOO} we consider three-body $B$ decays and showcase additional output-formatting options. Finally, in Section~\ref{sec:mass sum rules} we illustrate an application beyond decay amplitudes by deriving baryon mass sum rules, and we show the use of custom labels. The full code and outputs for all examples discussed in this section are included in the example tutorial notebook in the \href{https://github.com/Flavor-Sum-Rules/FlaSR}{GitHub repository}.

\subsection{$D^0 \to P^+ P^-$}
\label{sec:D to PP}

We start by demonstrating the physical mode of \verb|FlaSR| for the $U$-spin system of charm decays, $D^0 \to P^+ P^-$, where the $U$-spin multiplets $D^0$, $P^+$, and $P^-$ are defined in Eq.~\eqref{eq:DPPmultiplet}. This phenomenologically rich system has been studied extensively \cite{Brod:2012ud, Muller:2015lua, Grossman:2006jg, Hiller:2012xm, Grossman:2013lya, Grossman:2019xcj, gavrilova2022structure, Gavrilova:2026ryc, grossman2013su3}, including in the context of higher-order flavor sum rules \cite{gavrilova2022structure, grossman2013su3,Gavrilova:2026ryc, gavrilova2024direct}.

Neglecting corrections of $\mathcal{O}(\lambda^4)$, where $\lambda \approx 0.22$ is the Wolfenstein parameter, the effective Hamiltonian for fully hadronic charm decays transforms as a \U-spin triplet,
\begin{equation}
    \mathcal{H}_{\text{eff}}^{(0)} = \sum_{m=-1}^1 f_{1,m} H_m^1
    \label{eq:DPPHeff}
\, ,\end{equation}
where
\begin{equation}\label{eq:DPPHop}
    H_1^1 = (\overline{u}s)(\overline{d}c)\,, \qquad
    H_0^1 = \frac{(\overline{u}s)(\overline{s}c) - (\overline{u}d)(\overline{d}c)}{\sqrt{2}} \,, \qquad
    H_{-1}^1 = -(\overline{u}d)(\overline{s}c)\,,
\end{equation}
and
\begin{equation}
    f_{1,1} = V_{cd}^* V_{us} \approx -\lambda^2\,, \qquad
    f_{1,0} = \frac{V_{cs}^* V_{us} - V_{cd}^* V_{ud}}{\sqrt{2}} \approx \sqrt{2}\lambda\,, \qquad
    f_{1,-1} = -V_{cs}^* V_{ud} \approx -1
    \label{eq:DPPCKM}
\, ,\end{equation}
where, for simplicity of presentation in the code example below, the CKM factors are taken at leading order in $\lambda$. With this approximation, the system satisfies the assumptions summarized in Section~\ref{sec:scope}: all external states and the symmetry-limit Hamiltonian are given by irreducible \U-spin representations, and the full symmetry-related set of channels is kinematically allowed. Thus \texttt{FlaSR} can be used to derive amplitude and amplitude-squared sum rules for this system.

In \texttt{FlaSR}, particle multiplets are defined as ordered lists of the physical particles in the multiplet, while the Hamiltonian is specified as the ordered list of CKM factors associated with the components of the symmetry-limit Hamiltonian $\mathcal{H}_{\text{eff}}^{(0)}$. Throughout, we order the entries within each multiplet from highest to lowest $m$-QN:
\begin{Verbatim}[frame=single,commandchars=\\\{\},codes={\catcode`$=3\catcode`^=7\catcode`_=8}]
D0 = \{"$\texttt{D}^0$"\};
Pp = \{"$\texttt{K}^+$","$\pi^+$"\};
Pm = \{"$\pi^-$","$\texttt{K}^-$"\};
H = \{$-\lambda^2$,$\sqrt{2}*\lambda$,$-1$\};
\end{Verbatim}
Here, \verb|D0|, \verb|Pp|, and \verb|Pm| respectively define the particle multiplets $D^0$, $P^+$, and $P^-$ given in Eq.~\eqref{eq:DPPmultiplet}, while \verb|H| specifies the triplet Hamiltonian in Eqs.~\eqref{eq:DPPHeff}--\eqref{eq:DPPCKM}. Particle names are entered as strings, while the entries of \verb|H| can be either numbers or symbolic expressions. Note that for physical systems, all multiplets, including \U-spin singlets, must be defined explicitly.

To generate the sum rules, we use the \verb|generateSRs| function in the physical input mode (\verb|phys -> True|) with the physical multiplets \verb|D0|, \verb|Pp|, \verb|Pm| and the Hamiltonian \verb|H| defined above as arguments:
\begin{Verbatim}[frame=single,commandchars=\\\{\},codes={\catcode`$=3\catcode`^=7\catcode`_=8}]
system = generateSRs[\{D0\}, \{H\}, \{Pp, Pm\}, phys -> True];
\end{Verbatim}
Running \texttt{generateSRs} returns a Mathematica association that contains the full list of amplitudes of the system, along with the corresponding amplitude and amplitude-squared sum rules; see Table~\ref{tab:generateSRs output} for details.

Next, we print the results using the \verb|printSystem| function. For our display options, we choose to write the sum rules using $A$-type amplitudes (\verb|ampType| and \verb|amp2Type -> {A}|) in expanded form (\verb|expandSRs -> True|) and with their respective CKM factors restored (\verb|CKM -> True|):
\begin{Verbatim}[frame=single,commandchars=\\\{\},codes={\catcode`$=3\catcode`^=7},breaklines=true]
printSystem[system, ampType -> \{A\}, amp2Type -> \{A\}, expandSRs -> True, CKM -> True];
\end{Verbatim}

The full program output is given below:
\begin{Verbatim}[frame=single,commandchars=\\\{\},codes={\catcode`$=3\catcode`^=7\catcode`_=8\catcode`&=4}]
System: \{0,$\frac{1}{2}$,$\frac{1}{2}$,1\}
-------------------------
Number of would-be doublets: 4
In: \{0\}
H: \{1\}
Out: \{$\frac{1}{2}$,$\frac{1}{2}$\}
-------------------------
Amplitude table
Number of amplitudes: 4
a/s definitions: $\texttt{a}_\texttt{i} = \texttt{A}_\texttt{i} - \overline{\texttt{A}}_\texttt{i}$, $\texttt{s}_\texttt{i} = \texttt{A}_\texttt{i} + \overline{\texttt{A}}_\texttt{i}$
$\Delta$/$\Sigma$ definitions: $\Delta_\texttt{i} = |\texttt{A}_\texttt{i}|^2 - |\overline{\texttt{A}}_\texttt{i}|^2$, $\Sigma_\texttt{i} = |\texttt{A}_\texttt{i}|^2 + |\overline{\texttt{A}}_\texttt{i}|^2$
$\begin{matrix*}[l]\texttt{Process} && \texttt{QN label} && \texttt{n-tuple} && \texttt{Coord} && \texttt{CKM}\\\texttt{D}^0 \to \pi^+ \texttt{K}^- && 0 \overset{-1}{\to} -\frac{1}{2} -\frac{1}{2} && \texttt{(-,-,++)} && \texttt{(1)} && -1\\\texttt{D}^0 \to \texttt{K}^+ \pi^- && 0 \overset{+1}{\to} +\frac{1}{2} +\frac{1}{2} && \texttt{(+,+,--)} && && -\lambda^2\\\\\texttt{D}^0 \to \pi^+ \pi^- && 0 \overset{0}{\to} -\frac{1}{2} +\frac{1}{2} && \texttt{(-,+,-+)} && \texttt{(2)} && \sqrt{2}\lambda\\\texttt{D}^0 \to \texttt{K}^+ \texttt{K}^- && 0 \overset{0}{\to} +\frac{1}{2} -\frac{1}{2} && \texttt{(+,-,-+)} && && \sqrt{2}\lambda\end{matrix*}$
-------------------------
Amplitude sum rules
b = 0
Number of SRs: 2
$\begin{pmatrix}\frac{\texttt{A}(\texttt{D}^0 \to \texttt{K}^+ \pi^-)}{\lambda^2} - \texttt{A}(\texttt{D}^0 \to \pi^+ \texttt{K}^-)\\\frac{\texttt{A}(\texttt{D}^0 \to \pi^+ \pi^-)}{\sqrt{2}\lambda} - \frac{\texttt{A}(\texttt{D}^0 \to \texttt{K}^+ \texttt{K}^-)}{\sqrt{2}\lambda}\end{pmatrix}$
b = 1
Number of SRs: 1
$\begin{pmatrix}-\frac{\texttt{A}(\texttt{D}^0 \to \texttt{K}^+ \texttt{K}^-)}{\lambda} - \frac{\texttt{A}(\texttt{D}^0 \to \texttt{K}^+ \pi^-)}{\lambda^2} - \texttt{A}(\texttt{D}^0 \to \pi^+ \texttt{K}^-) - \frac{\texttt{A}(\texttt{D}^0 \to \pi^+ \pi^-)}{\lambda}\end{pmatrix}$
-------------------------
Amplitude-squared sum rules
b = 0
Number of SRs: 2
$\begin{pmatrix}|\texttt{A}(\texttt{D}^0 \to \pi^+ \texttt{K}^-)|^2 - \left|\frac{\texttt{A}(\texttt{D}^0 \to \texttt{K}^+ \pi^-)}{\lambda^2}\right|^2\\\frac{1}{2}\left|\frac{\texttt{A}(\texttt{D}^0 \to \pi^+ \pi^-)}{\lambda}\right|^2 - \frac{1}{2}\left|\frac{\texttt{A}(\texttt{D}^0 \to \texttt{K}^+ \texttt{K}^-)}{\lambda}\right|^2\end{pmatrix}$
b = 1
Number of SRs: 1
$\begin{pmatrix}-\left|\frac{\texttt{A}(\texttt{D}^0 \to \texttt{K}^+ \texttt{K}^-)}{\lambda}\right|^2 + \left|\frac{\texttt{A}(\texttt{D}^0 \to \texttt{K}^+ \pi^-)}{\lambda^2}\right|^2 + |\texttt{A}(\texttt{D}^0 \to \pi^+ \texttt{K}^-)|^2 - \left|\frac{\texttt{A}(\texttt{D}^0 \to \pi^+ \pi^-)}{\lambda}\right|^2\end{pmatrix}$
\end{Verbatim}

Since the $D^0 \to P^+P^-$ system considered here has exactly the same group-theoretic structure as the abstract system of two doublets and a triplet discussed in Section~\ref{sec:instructions}, it is instructive to compare the two outputs. Compared to the output for the abstract group-theoretic system in Section~\ref{sec:instructions}, the amplitude table for the physical $D^0 \to P^+ P^-$ system contains by default two additional columns: one listing the physical processes in terms of the initial- and final-state particles, and one listing the CKM factors specified when defining \texttt{H}, with the appropriate CKM factor automatically matched to each amplitude. Since the CKM factors have been restored in the sum rule output, the amplitudes should be interpreted as physical amplitudes $\mathcal{A}_i$. In addition, the default behavior in the physical mode is to present the sum rules in terms of amplitudes labeled by their corresponding physical processes. The structure of the sum rules remains the same; see Eqs.~\eqref{eq:instructions asrb0}--\eqref{eq:instructions a2srb1}. The \texttt{FlaSR} output reproduces known amplitude sum rules, see, for example, Eqs.~(5.24)--(5.25) of Ref.~\cite{gavrilova2022structure}, as well as known amplitude-squared sum rules, see, for example, the third equation in Section~3.4 of Ref.~\cite{grossman2013su3}.

\subsection{$D_q^- \to P^+ P^- P^-$}
\label{sec:DPPP}

Next we consider a system of three-body charm decays, $D_q^- \to P^+ P^- P^-$, with two identical $P^-$ doublets in the final state. The final-state doublets are defined in Eq.~\eqref{eq:DPPmultiplet}, $D_q^-$ is given by
\begin{equation}\label{eq:D_ds_def}
D_q^-\equiv
\begin{bmatrix}
D^-\\[2pt]
D_s^-
\end{bmatrix}
=
\begin{bmatrix}
\ket{d\overline{c}}\\[2pt]
\ket{s\overline{c}}
\end{bmatrix}\,,
\end{equation}
and the charm Hamiltonian is defined in Eqs.~\eqref{eq:DPPHeff}--\eqref{eq:DPPCKM}. In what follows, we use \texttt{FlaSR} to derive amplitude-squared sum rules both in the differential case (\texttt{obs -> "Diff"}) and in the integrated case (\texttt{obs -> "Int"}), and we comment on the differences in the outputs and their interpretations.

We start by defining the physical multiplets in the initial and final state of the $D_q^-\to P^+P^-P^-$ system and the Hamiltonian:
\begin{Verbatim}[frame=single,commandchars=\\\{\},codes={\catcode`$=3\catcode`^=7\catcode`_=8}]
Dq = \{"$\texttt{D}^-$","$\texttt{D}_\texttt{s}^-$"\};
Pp = \{"$\texttt{K}^+$","$\pi^+$"\};
Pm = \{"$\pi^-$","$\texttt{K}^-$"\};
H = \{$-\lambda^2$,$\sqrt{2}*\lambda$,$-1$\};
\end{Verbatim}
Here, \verb|Dq|, \verb|Pp|, and \verb|Pm| define the particle multiplets $D_q^-$, $P^+$, and $P^-$, respectively, see Eqs.~\eqref{eq:DPPmultiplet} and~\eqref{eq:D_ds_def}, and \verb|H| is the triplet Hamiltonian, see Eqs.~\eqref{eq:DPPHeff}--\eqref{eq:DPPCKM}.

Next, we run the \verb|generateSRs| function in the physical input mode (\verb|phys -> True|) with the physical multiplets \verb|Dq|, \verb|Pm|, \verb|Pp| and the Hamiltonian \verb|H| as arguments. Here we choose the default differential setting \texttt{obs -> "Diff"}, which we do not need to specify explicitly:
\begin{Verbatim}[frame=single,commandchars=\\\{\},codes={\catcode`$=3\catcode`^=7\catcode`_=8}]
system = generateSRs[\{Dq\}, \{H\}, \{Pp, Pm, Pm\}, phys -> True];
\end{Verbatim}

We pass the resulting association, \texttt{system}, to \texttt{printSystem}. For the display options, we suppress the amplitude sum rules (\verb|showASRs -> False|), so there is no need to specify \verb|ampType|. We choose to display the amplitude-squared sum rules in terms of squared CKM-free (the default, \verb|CKM -> False|) $A$-type amplitudes (\verb|amp2Type -> {A}|) and in expanded form (\verb|expandSRs -> True|):
\begin{Verbatim}[frame=single,commandchars=\\\{\},codes={\catcode`$=3\catcode`^=7},breaklines=true]
printSystem[system, showASRs -> False, amp2Type -> \{A\}, expandSRs -> True];
\end{Verbatim}

The resulting output is shown below:
\begin{Verbatim}[frame=single,commandchars=\\\{\},codes={\catcode`$=3\catcode`^=7\catcode`_=8\catcode`&=4}]
System: \{$\frac{1}{2}$,$\frac{1}{2}$,$\frac{1}{2}$,$\frac{1}{2}$,1\}
-------------------------
Number of would-be doublets: 6
In: \{$\frac{1}{2}$\}
H: \{1\}
Out: \{$\frac{1}{2}$,$\frac{1}{2}$,$\frac{1}{2}$\}
-------------------------
Amplitude table
Number of amplitudes: 14
a/s definitions: $\texttt{a}_\texttt{i} = \texttt{A}_\texttt{i} - \overline{\texttt{A}}_\texttt{i}$, $\texttt{s}_\texttt{i} = \texttt{A}_\texttt{i} + \overline{\texttt{A}}_\texttt{i}$
$\Delta$/$\Sigma$ definitions: $\Delta_\texttt{i} = |\texttt{A}_\texttt{i}|^2 - |\overline{\texttt{A}}_\texttt{i}|^2$, $\Sigma_\texttt{i} = |\texttt{A}_\texttt{i}|^2 + |\overline{\texttt{A}}_\texttt{i}|^2$
$\begin{matrix*}[l]\texttt{Process} && \texttt{QN label} && \texttt{n-tuple} && \texttt{Coord} && \texttt{CKM}\\\texttt{D}^- \to \pi^+ \texttt{K}^- \pi^- && +\frac{1}{2} \overset{-1}{\to} -\frac{1}{2} -\frac{1}{2} +\frac{1}{2} && \texttt{(-,-,-,+,++)} && \texttt{(1,2)} && -1\\\texttt{D}_\texttt{s}^- \to \texttt{K}^+ \pi^- \texttt{K}^- && -\frac{1}{2} \overset{+1}{\to} +\frac{1}{2} +\frac{1}{2} -\frac{1}{2} && \texttt{(+,+,+,-,--)} && && -\lambda^2\\\\\texttt{D}^- \to \pi^+ \pi^- \texttt{K}^- && +\frac{1}{2} \overset{-1}{\to} -\frac{1}{2} +\frac{1}{2} -\frac{1}{2} && \texttt{(-,-,+,-,++)} && \texttt{(1,3)} && -1\\\texttt{D}_\texttt{s}^- \to \texttt{K}^+ \texttt{K}^- \pi^- && -\frac{1}{2} \overset{+1}{\to} +\frac{1}{2} -\frac{1}{2} +\frac{1}{2} && \texttt{(+,+,-,+,--)} && && -\lambda^2\\\\\texttt{D}^- \to \pi^+ \pi^- \pi^- && +\frac{1}{2} \overset{0}{\to} -\frac{1}{2} +\frac{1}{2} +\frac{1}{2} && \texttt{(-,-,+,+,-+)} && \texttt{(1,4)} && \sqrt{2}\lambda\\\texttt{D}_\texttt{s}^- \to \texttt{K}^+ \texttt{K}^- \texttt{K}^- && -\frac{1}{2} \overset{0}{\to} +\frac{1}{2} -\frac{1}{2} -\frac{1}{2} && \texttt{(+,+,-,-,-+)} && && \sqrt{2}\lambda\\\\\texttt{D}^- \to \texttt{K}^+ \texttt{K}^- \texttt{K}^- && +\frac{1}{2} \overset{-1}{\to} +\frac{1}{2} -\frac{1}{2} -\frac{1}{2} && \texttt{(-,+,-,-,++)} && \texttt{(2,3)} && -1\\\texttt{D}_\texttt{s}^- \to \pi^+ \pi^- \pi^- && -\frac{1}{2} \overset{+1}{\to} -\frac{1}{2} +\frac{1}{2} +\frac{1}{2} && \texttt{(+,-,+,+,--)} && && -\lambda^2\\\\\texttt{D}^- \to \texttt{K}^+ \texttt{K}^- \pi^- && +\frac{1}{2} \overset{0}{\to} +\frac{1}{2} -\frac{1}{2} +\frac{1}{2} && \texttt{(-,+,-,+,-+)} && \texttt{(2,4)} && \sqrt{2}\lambda\\\texttt{D}_\texttt{s}^- \to \pi^+ \pi^- \texttt{K}^- && -\frac{1}{2} \overset{0}{\to} -\frac{1}{2} +\frac{1}{2} -\frac{1}{2} && \texttt{(+,-,+,-,-+)} && && \sqrt{2}\lambda\\\\\texttt{D}^- \to \texttt{K}^+ \pi^- \texttt{K}^- && +\frac{1}{2} \overset{0}{\to} +\frac{1}{2} +\frac{1}{2} -\frac{1}{2} && \texttt{(-,+,+,-,-+)} && \texttt{(3,4)} && \sqrt{2}\lambda\\\texttt{D}_\texttt{s}^- \to \pi^+ \texttt{K}^- \pi^- && -\frac{1}{2} \overset{0}{\to} -\frac{1}{2} -\frac{1}{2} +\frac{1}{2} && \texttt{(+,-,-,+,-+)} && && \sqrt{2}\lambda\\\\\texttt{D}^- \to \texttt{K}^+ \pi^- \pi^- && +\frac{1}{2} \overset{+1}{\to} +\frac{1}{2} +\frac{1}{2} +\frac{1}{2} && \texttt{(-,+,+,+,--)} && \texttt{(4,4)} && -\lambda^2\\\texttt{D}_\texttt{s}^- \to \pi^+ \texttt{K}^- \texttt{K}^- && -\frac{1}{2} \overset{-1}{\to} -\frac{1}{2} -\frac{1}{2} -\frac{1}{2} && \texttt{(+,-,-,-,++)} && && -1\\\end{matrix*}$
-------------------------
Amplitude-squared sum rules
b = 0
Number of SRs: 7
$\begin{pmatrix}|\texttt{A}(\texttt{D}^- \to \pi^+ \texttt{K}^- \pi^-)|^2 - |\texttt{A}(\texttt{D}_\texttt{s}^- \to \texttt{K}^+ \pi^- \texttt{K}^-)|^2\\|\texttt{A}(\texttt{D}^- \to \pi^+ \pi^- \texttt{K}^-)|^2 - |\texttt{A}(\texttt{D}_\texttt{s}^- \to \texttt{K}^+ \texttt{K}^- \pi^-)|^2\\|\texttt{A}(\texttt{D}^- \to \pi^+ \pi^- \pi^-)|^2 - |\texttt{A}(\texttt{D}_\texttt{s}^- \to \texttt{K}^+ \texttt{K}^- \texttt{K}^-)|^2\\|\texttt{A}(\texttt{D}^- \to \texttt{K}^+ \texttt{K}^- \texttt{K}^-)|^2 - |\texttt{A}(\texttt{D}_\texttt{s}^- \to \pi^+ \pi^- \pi^-)|^2\\|\texttt{A}(\texttt{D}^- \to \texttt{K}^+ \texttt{K}^- \pi^-)|^2 - |\texttt{A}(\texttt{D}_\texttt{s}^- \to \pi^+ \pi^- \texttt{K}^-)|^2\\|\texttt{A}(\texttt{D}^- \to \texttt{K}^+ \pi^- \texttt{K}^-)|^2 - |\texttt{A}(\texttt{D}_\texttt{s}^- \to \pi^+ \texttt{K}^- \pi^-)|^2\\|\texttt{A}(\texttt{D}^- \to \texttt{K}^+ \pi^- \pi^-)|^2 - |\texttt{A}(\texttt{D}_\texttt{s}^- \to \pi^+ \texttt{K}^- \texttt{K}^-)|^2\end{pmatrix}$
b = 1
Number of SRs: 1
$\begin{pmatrix*}[l]|\texttt{A}(\texttt{D}^- \to \texttt{K}^+ \texttt{K}^- \texttt{K}^-)|^2 - 2|\texttt{A}(\texttt{D}^- \to \texttt{K}^+ \texttt{K}^- \pi^-)|^2 - 2|\texttt{A}(\texttt{D}^- \to \texttt{K}^+ \pi^- \texttt{K}^-)|^2 + |\texttt{A}(\texttt{D}^- \to \texttt{K}^+ \pi^- \pi^-)|^2\\\quad+|\texttt{A}(\texttt{D}^- \to \pi^+ \texttt{K}^- \pi^-)|^2 + |\texttt{A}(\texttt{D}^- \to \pi^+ \pi^- \texttt{K}^-)|^2 - 2|\texttt{A}(\texttt{D}^- \to \pi^+ \pi^- \pi^-)|^2 - 2|\texttt{A}(\texttt{D}_\texttt{s}^- \to \texttt{K}^+ \texttt{K}^- \texttt{K}^-)|^2\\\quad+|\texttt{A}(\texttt{D}_\texttt{s}^- \to \texttt{K}^+ \texttt{K}^- \pi^-)|^2 + |\texttt{A}(\texttt{D}_\texttt{s}^- \to \texttt{K}^+ \pi^- \texttt{K}^-)|^2 + |\texttt{A}(\texttt{D}_\texttt{s}^- \to \pi^+ \texttt{K}^- \texttt{K}^-)|^2 - 2|\texttt{A}(\texttt{D}_\texttt{s}^- \to \pi^+ \texttt{K}^- \pi^-)|^2\\\quad-2|\texttt{A}(\texttt{D}_\texttt{s}^- \to \pi^+ \pi^- \texttt{K}^-)|^2 + |\texttt{A}(\texttt{D}_\texttt{s}^- \to \pi^+ \pi^- \pi^-)|^2\end{pmatrix*}$
b = 2
Number of SRs: 0
No sum rules found at this order.
\end{Verbatim}

We emphasize the following about the interpretation of the output in the differential setting (\texttt{obs -> "Diff"}):
\begin{itemize}
    \item The amplitudes are always understood as momentum-dependent quantities. Accordingly, different positions in the final-state correspond to different implicit momentum assignments. This is why the amplitude table above contains, for example, both amplitudes $A(D^- \to \pi^+ K^- \pi^-)$ and $A(D^- \to \pi^+ \pi^- K^-)$: they correspond to different momentum-labeled configurations and are therefore distinct physical amplitudes. Consequently, if one were able to test the amplitude sum rules directly (they are suppressed in the above output due to \verb|showASRs -> False|), one would have to include all the amplitudes listed, including those whose labels differ only by a permutation of the final-state particles.
    
    \item In the differential mode, both the amplitude sum rules and the amplitude-squared sum rules are written in terms of the full set of amplitudes in the system, including those whose labels differ only by a permutation of the final-state particles. The amplitude-squared sum rules output can be directly interpreted as relations among differential observables proportional to $|A|^2$, in this case CKM-normalized differential decay rates. At order $b=0$ this identification is immediate, since in the \U-spin symmetry limit all channels share the same phase-space factors. For the $b=1$ sum rule, by the symmetry argument of Section~2.1 of Ref.~\cite{Gavrilova:2026ryc}, this interpretation remains valid even with phase-space differences taken into account.
\end{itemize}

To convert the amplitude-squared sum rules in the output above into relations among integrated observables, which are integrated decay rates in the example we consider, we run \verb|generateSRs| with the setting \verb|obs -> "Int"|:
\begin{Verbatim}[frame=single,commandchars=\\\{\},codes={\catcode`$=3\catcode`^=7\catcode`_=8}]
system = generateSRs[\{Dq\}, \{H\}, \{Pp, Pm, Pm\}, phys -> True, obs -> "Int"];
\end{Verbatim}

We run \texttt{printSystem} with the same settings as before, along with an additional setting to suppress the system summary (\texttt{showReps -> False}), which is the same in both cases. We obtain the following output:
\begin{Verbatim}[frame=single,commandchars=\\\{\},codes={\catcode`$=3\catcode`^=7\catcode`_=8\catcode`&=4}]
Amplitude table
Number of amplitudes: 14
a/s definitions: $\texttt{a}_\texttt{i} = \texttt{A}_\texttt{i} - \overline{\texttt{A}}_\texttt{i}$, $\texttt{s}_\texttt{i} = \texttt{A}_\texttt{i} + \overline{\texttt{A}}_\texttt{i}$
$\Delta$/$\Sigma$ definitions: $\Delta_\texttt{i} = |\texttt{A}_\texttt{i}|^2 - |\overline{\texttt{A}}_\texttt{i}|^2$, $\Sigma_\texttt{i} = |\texttt{A}_\texttt{i}|^2 + |\overline{\texttt{A}}_\texttt{i}|^2$
$\begin{matrix*}[l]\texttt{Process} && \texttt{QN label} && \texttt{n-tuple} && \texttt{Coord} && \texttt{CKM} && \texttt{Integrated channel ID}\\\texttt{D}^- \to \pi^+ \texttt{K}^- \pi^- && +\frac{1}{2} \overset{-1}{\to} -\frac{1}{2} -\frac{1}{2} +\frac{1}{2} && \texttt{(-,-,-,+,++)} && \texttt{(1,2)} && -1 && 1\\\texttt{D}_\texttt{s}^- \to \texttt{K}^+ \pi^- \texttt{K}^- && -\frac{1}{2} \overset{+1}{\to} +\frac{1}{2} +\frac{1}{2} -\frac{1}{2} && \texttt{(+,+,+,-,--)} && && -\lambda^2 && 2\\\\\texttt{D}^- \to \pi^+ \pi^- \texttt{K}^- && +\frac{1}{2} \overset{-1}{\to} -\frac{1}{2} +\frac{1}{2} -\frac{1}{2} && \texttt{(-,-,+,-,++)} && \texttt{(1,3)} && -1 && 1\\\texttt{D}_\texttt{s}^- \to \texttt{K}^+ \texttt{K}^- \pi^- && -\frac{1}{2} \overset{+1}{\to} +\frac{1}{2} -\frac{1}{2} +\frac{1}{2} && \texttt{(+,+,-,+,--)} && && -\lambda^2 && 2\\\\\texttt{D}^- \to \pi^+ \pi^- \pi^- && +\frac{1}{2} \overset{0}{\to} -\frac{1}{2} +\frac{1}{2} +\frac{1}{2} && \texttt{(-,-,+,+,-+)} && \texttt{(1,4)} && \sqrt{2}\lambda && 3\\\texttt{D}_\texttt{s}^- \to \texttt{K}^+ \texttt{K}^- \texttt{K}^- && -\frac{1}{2} \overset{0}{\to} +\frac{1}{2} -\frac{1}{2} -\frac{1}{2} && \texttt{(+,+,-,-,-+)} && && \sqrt{2}\lambda && 4\\\\\texttt{D}^- \to \texttt{K}^+ \texttt{K}^- \texttt{K}^- && +\frac{1}{2} \overset{-1}{\to} +\frac{1}{2} -\frac{1}{2} -\frac{1}{2} && \texttt{(-,+,-,-,++)} && \texttt{(2,3)} && -1 && 5\\\texttt{D}_\texttt{s}^- \to \pi^+ \pi^- \pi^- && -\frac{1}{2} \overset{+1}{\to} -\frac{1}{2} +\frac{1}{2} +\frac{1}{2} && \texttt{(+,-,+,+,--)} && && -\lambda^2 && 6\\\\\texttt{D}^- \to \texttt{K}^+ \texttt{K}^- \pi^- && +\frac{1}{2} \overset{0}{\to} +\frac{1}{2} -\frac{1}{2} +\frac{1}{2} && \texttt{(-,+,-,+,-+)} && \texttt{(2,4)} && \sqrt{2}\lambda && 7\\\texttt{D}_\texttt{s}^- \to \pi^+ \pi^- \texttt{K}^- && -\frac{1}{2} \overset{0}{\to} -\frac{1}{2} +\frac{1}{2} -\frac{1}{2} && \texttt{(+,-,+,-,-+)} && && \sqrt{2}\lambda && 8\\\end{matrix*}$
$\begin{matrix*}[l]\texttt{D}^- \to \texttt{K}^+ \pi^- \texttt{K}^- && +\frac{1}{2} \overset{0}{\to} +\frac{1}{2} +\frac{1}{2} -\frac{1}{2} && \texttt{(-,+,+,-,-+)} && \texttt{(3,4)} && \sqrt{2}\lambda && 7\\\texttt{D}_\texttt{s}^- \to \pi^+ \texttt{K}^- \pi^- && -\frac{1}{2} \overset{0}{\to} -\frac{1}{2} -\frac{1}{2} +\frac{1}{2} && \texttt{(+,-,-,+,-+)} && && \sqrt{2}\lambda && 8\\\\\texttt{D}^- \to \texttt{K}^+ \pi^- \pi^- && +\frac{1}{2} \overset{+1}{\to} +\frac{1}{2} +\frac{1}{2} +\frac{1}{2} && \texttt{(-,+,+,+,--)} && \texttt{(4,4)} && -\lambda^2 && 9\\\texttt{D}_\texttt{s}^- \to \pi^+ \texttt{K}^- \texttt{K}^- && -\frac{1}{2} \overset{-1}{\to} -\frac{1}{2} -\frac{1}{2} -\frac{1}{2} && \texttt{(+,-,-,-,++)} && && -1 && 10\\\end{matrix*}$
-------------------------
Amplitude-squared sum rules
b = 0
Number of SRs: 5
$\begin{pmatrix}|\texttt{A}(\texttt{D}^- \to \pi^+ \texttt{K}^- \pi^-)|^2 - |\texttt{A}(\texttt{D}_\texttt{s}^- \to \texttt{K}^+ \pi^- \texttt{K}^-)|^2\\|\texttt{A}(\texttt{D}^- \to \pi^+ \pi^- \pi^-)|^2 - |\texttt{A}(\texttt{D}_\texttt{s}^- \to \texttt{K}^+ \texttt{K}^- \texttt{K}^-)|^2\\|\texttt{A}(\texttt{D}^- \to \texttt{K}^+ \texttt{K}^- \texttt{K}^-)|^2 - |\texttt{A}(\texttt{D}_\texttt{s}^- \to \pi^+ \pi^- \pi^-)|^2\\|\texttt{A}(\texttt{D}^- \to \texttt{K}^+ \texttt{K}^- \pi^-)|^2 - |\texttt{A}(\texttt{D}_\texttt{s}^- \to \pi^+ \pi^- \texttt{K}^-)|^2\\|\texttt{A}(\texttt{D}^- \to \texttt{K}^+ \pi^- \pi^-)|^2 - |\texttt{A}(\texttt{D}_\texttt{s}^- \to \pi^+ \texttt{K}^- \texttt{K}^-)|^2\end{pmatrix}$
b = 1
Number of SRs: 1
$\begin{pmatrix*}[l]|\texttt{A}(\texttt{D}^- \to \texttt{K}^+ \texttt{K}^- \texttt{K}^-)|^2 - 2|\texttt{A}(\texttt{D}^- \to \texttt{K}^+ \texttt{K}^- \pi^-)|^2 + |\texttt{A}(\texttt{D}^- \to \texttt{K}^+ \pi^- \pi^-)|^2 + |\texttt{A}(\texttt{D}^- \to \pi^+ \texttt{K}^- \pi^-)|^2\\\quad-2|\texttt{A}(\texttt{D}^- \to \pi^+ \pi^- \pi^-)|^2 - 2|\texttt{A}(\texttt{D}_\texttt{s}^- \to \texttt{K}^+ \texttt{K}^- \texttt{K}^-)|^2 + |\texttt{A}(\texttt{D}_\texttt{s}^- \to \texttt{K}^+ \pi^- \texttt{K}^-)|^2 + |\texttt{A}(\texttt{D}_\texttt{s}^- \to \pi^+ \texttt{K}^- \texttt{K}^-)|^2\\\quad-2|\texttt{A}(\texttt{D}_\texttt{s}^- \to \pi^+ \pi^- \texttt{K}^-)|^2 + |\texttt{A}(\texttt{D}_\texttt{s}^- \to \pi^+ \pi^- \pi^-)|^2\end{pmatrix*}$
b = 2
Number of SRs: 0
No sum rules found at this order.
\end{Verbatim}

In the integrated setting (\texttt{obs -> "Int"}), the amplitude-squared output should be understood to have been obtained by taking the corresponding differential output and integrating it over the final-state phase space. Under this integration, channels that differ only by permutations of momenta of the final state particles are identified together. This is reflected by the addition of a new column in the amplitude table, \verb|Integrated channel ID|, which enumerates the unique integrated channels by their order of appearance in the table. Processes with the same number correspond to the same integrated channel ID. For example, the two differential amplitudes $A(D^- \to \pi^+ K^- \pi^-)$ and $A(D^- \to \pi^+ \pi^- K^-)$ are combined into integrated channel 1, $A(D^- \to \pi^+ K^- \pi^-)$. Additionally, the usual combinatorial factors arising from the definition of the phase-space integral for final states with identical particles are included automatically by \texttt{FlaSR}. The amplitude-squared sum rules in the output for \texttt{obs -> "Int"} should therefore be interpreted as relations between quantities of the form
\begin{equation}
    \int d\Pi \,|A|^2 \,,
\end{equation}
which, in the case considered here, are CKM-normalized partial decay rates. At order $b=0$ this identification is immediate. For the $b=1$ sum rules, the same interpretation remains by the symmetry argument of Section~2.1 of Ref.~\cite{Gavrilova:2026ryc}.

The \texttt{FlaSR} output in this subsection reproduces the rate sum rules derived in Ref.~\cite{Gavrilova:2026ryc} up to factors of $2$ accounted for by CKM factors; see Section~4.3 and Appendix~B therein.

\subsection{$\overline{B}^0 \to D^0 \Omega_B^- \overline{\Omega}_B^+$}
\label{sec:B to DOO}

We now consider a $U$-spin system of three-body $B$-decays, $\overline{B}^0 \to D^0 \Omega_B^- \overline{\Omega}_B^+$, where
\begin{gather}
\overline{B}^0 =
\begin{bmatrix}
    \overline{B}_s^0 \\
    \overline{B}_d^0
\end{bmatrix}
=
\begin{bmatrix}
    |b\overline{s}\rangle \\
    -|b\overline{d}\rangle
\end{bmatrix}\,, \nonumber\\
\Omega_B^- =
\begin{bmatrix}
    \Delta^- \\
    \Sigma^{*-} \\
    \Xi^{*-} \\
    \Omega^-
\end{bmatrix}
=
\begin{bmatrix}
    |ddd\rangle \\
    |dds\rangle \\
    |dss\rangle \\
    |sss\rangle
\end{bmatrix}\,, \quad
\overline{\Omega}_B^+ =
\begin{bmatrix}
    \overline{\Omega}^+ \\
    \overline{\Xi}^{*+} \\
    \overline{\Sigma}^{*+} \\
    \overline{\Delta}^+
\end{bmatrix}
=
\begin{bmatrix}
    |\overline{sss}\rangle \\
    -|\overline{dss}\rangle \\
    |\overline{dds}\rangle \\
    -|\overline{ddd}\rangle
\end{bmatrix}
\label{eq:BDOOmultiplet}
\,,\end{gather}
and where $D^0$ is a $U$-spin singlet defined in Eq.~\eqref{eq:DPPmultiplet}. In the \U-spin limit, the effective Hamiltonian governing these decays is given by a doublet,
\begin{equation}
    \mathcal{H}_{\text{eff}}^{(0)} = \sum_{m=-1/2}^{1/2} f_{1/2,m} H_m^{1/2}
\label{eq:BDOOHeff}
\,,\end{equation}
where
\begin{equation}
    H_{1/2}^{1/2} = (\overline{c}b)(\overline{s}u)\,, \quad
    H_{-1/2}^{1/2} = (\overline{c}b)(\overline{d}u)
\label{eq:BDOOHop}
\,,\end{equation}
and the CKM factors, using the Wolfenstein parametrization and keeping only the leading order in $\lambda$, are
\begin{equation}\label{eq:BDOOCKM}
    f_{1/2,1/2} = V_{cb} V_{us}^* \approx A\lambda^3\,, \qquad f_{1/2,-1/2} = V_{cb} V_{ud}^* \approx A\lambda^2\,.
\end{equation}

In \texttt{FlaSR}, we derive the sum rules by first defining the physical multiplets in the system. To prevent naming conflicts between the Wolfenstein parameter $A$ and the symbol for $A$-type amplitudes, we enter the Wolfenstein parameter as \verb|\[CapitalAlpha]| and reserve the regular \verb|A| for amplitudes. Both are rendered similarly, but the symbolic difference ensures that Mathematica treats them independently.
\begin{Verbatim}[frame=single,commandchars=\\\{\},codes={\catcode`$=3\catcode`^=7\catcode`_=8},breaklines=true]
B0bar = \{"$\overline{\texttt{B}}_\texttt{s}^0$","$\overline{\texttt{B}}_\texttt{d}^0$"\};
D0 = \{"$\texttt{D}^0$"\};
OBm = \{"$\Delta^-$","$\Sigma^{*-}$","$\Xi^{*-}$","$\Omega^{-}$"\};
OBpbar = \{"$\overline{\Omega}^{+}$","$\overline{\Xi}^{*+}$","$\overline{\Sigma}^{*+}$","$\overline{\Delta}^{+}$"\};
H = \{A$*\lambda^3$,A$*\lambda^2$\};
\end{Verbatim}

We generate the system, using the physical input mode (\verb|phys -> True|) and choosing to work with integrated observables (\verb|obs -> "Int"|)\footnote{Note that, for the $\overline{B}^0 \to D^0 \Omega_B^- \overline{\Omega}_B^+$ system, choosing either \texttt{obs -> "Int"} or \texttt{obs -> "Diff"} does not affect the result since there are no identical irreps in the final state.}:
\begin{Verbatim}[frame=single,commandchars=\\\{\},codes={\catcode`$=3\catcode`^=7\catcode`_=8},breaklines=true]
system = generateSRs[\{B0bar\}, \{H\}, \{D0, OBm, OBpbar\}, phys -> True, obs -> "Int"];
\end{Verbatim}

To write amplitude-squared sum rules in terms of symbols representing quantities proportional to $|A|^2$, we must indicate the quadratic dependence of the symbol assigned to \verb|amp2Type| through \verb|amp2Quad -> True|. For example, we are interested in rates, so we set \texttt{amp2Type -> \{$\Gamma$\}} and indicate that $\Gamma \propto |A|^2$ through \verb|amp2Quad -> True|. We choose to use the $A$-type amplitude basis for amplitude sum rules (\verb|ampType -> {A}|) and to present all sum rules in expanded form (\verb|expandSRs -> True|):
\begin{Verbatim}[frame=single,commandchars=\\\{\},codes={\catcode`$=3\catcode`^=7\catcode`_=8},breaklines=true]
printSystem[system, ampType -> \{A\}, amp2Type -> \{$\Gamma$\}, amp2Quad -> True, expandSRs -> True];
\end{Verbatim}

Below, we present a representative excerpt of the \texttt{FlaSR} output. To illustrate the capabilities of the package while keeping the output concise, we show only the higher-order rate sum rule output in full. Because we have implicitly set \verb|CKM -> False|, the CKM factors are omitted from display, and the sum rules should be interpreted as relations between CKM-normalized amplitudes and CKM-normalized rates, respectively.
\begin{Verbatim}[frame=single,commandchars=\\\{\},codes={\catcode`$=3\catcode`^=7\catcode`_=8\catcode`&=4}]
System: \{0,$\frac{1}{2}$,$\frac{1}{2}$,$\frac{3}{2}$,$\frac{3}{2}$\}
-------------------------
Number of would-be doublets: 8
In: \{$\frac{1}{2}$\}
H: \{$\frac{1}{2}$\}
Out: \{0,$\frac{3}{2}$,$\frac{3}{2}$\}
-------------------------
Amplitude table
Number of amplitudes: 14
a/s definitions: $\texttt{a}_\texttt{i} = \texttt{A}_\texttt{i} - \overline{\texttt{A}}_\texttt{i}$, $\texttt{s}_\texttt{i} = \texttt{A}_\texttt{i} + \overline{\texttt{A}}_\texttt{i}$
$\Delta$/$\Sigma$ definitions: $\Delta_\texttt{i} = |\texttt{A}_\texttt{i}|^2 - |\overline{\texttt{A}}_\texttt{i}|^2$, $\Sigma_\texttt{i} = |\texttt{A}_\texttt{i}|^2 + |\overline{\texttt{A}}_\texttt{i}|^2$
$\begin{matrix*}[l]\texttt{Process} && \texttt{QN label} && \texttt{n-tuple} && \texttt{Coord} && \texttt{CKM} && \texttt{Integrated channel ID}\\\overline{\texttt{B}}_\texttt{s}^0 \to \texttt{D}^0 \Xi^{*-} \overline{\Omega}^{+} && +\frac{1}{2} \overset{+\frac{1}{2}}{\to} 0 -\frac{1}{2} +\frac{3}{2} && \texttt{(-,-,--+,+++)} && \texttt{(1,2,2)} && \texttt{A}\lambda^3 && 1\\\overline{\texttt{B}}_\texttt{d}^0 \to \texttt{D}^0 \Sigma^{*-} \overline{\Delta}^{+} && -\frac{1}{2} \overset{-\frac{1}{2}}{\to} 0 +\frac{1}{2} -\frac{3}{2} && \texttt{(+,+,-++,---)} && && \texttt{A}\lambda^2 && 2\\\\\overline{\texttt{B}}_\texttt{s}^0 \to \texttt{D}^0 \Sigma^{*-} \overline{\Xi}^{*+} && +\frac{1}{2} \overset{+\frac{1}{2}}{\to} 0 +\frac{1}{2} +\frac{1}{2} && \texttt{(-,-,-++,-++)} && \texttt{(1,2,3)} && \texttt{A}\lambda^3 && 3\\\overline{\texttt{B}}_\texttt{d}^0 \to \texttt{D}^0 \Xi^{*-} \overline{\Sigma}^{*+} && -\frac{1}{2} \overset{-\frac{1}{2}}{\to} 0 -\frac{1}{2} -\frac{1}{2} && \texttt{(+,+,--+,--+)} && && \texttt{A}\lambda^2 && 4\\\\\overline{\texttt{B}}_\texttt{s}^0 \to \texttt{D}^0 \Delta^{-} \overline{\Sigma}^{*+} && +\frac{1}{2} \overset{+\frac{1}{2}}{\to} 0 +\frac{3}{2} -\frac{1}{2} && \texttt{(-,-,+++,--+)} && \texttt{(1,3,3)} && \texttt{A}\lambda^3 && 5\\\overline{\texttt{B}}_\texttt{d}^0 \to \texttt{D}^0 \Omega^{-} \overline{\Xi}^{*+} && -\frac{1}{2} \overset{-\frac{1}{2}}{\to} 0 -\frac{3}{2} +\frac{1}{2} && \texttt{(+,+,---,-++)} && && \texttt{A}\lambda^2 && 6\\\hfill\vdots\hfill && \hfill\vdots\hfill && \hfill\vdots\hfill && \hfill\vdots\hfill && \hfill\vdots\hfill && \hfill\vdots\hfill\end{matrix*}$
-------------------------
Amplitude sum rules
b = 0
Number of SRs: 7
$\begin{pmatrix*}[l]\texttt{A}(\overline{\texttt{B}}_\texttt{s}^0 \to \texttt{D}^0 \Xi^{*-} \overline{\Omega}^{+}) - \texttt{A}(\overline{\texttt{B}}_\texttt{d}^0 \to \texttt{D}^0 \Sigma^{*-} \overline{\Delta}^{+})\\\hfill\vdots\hfill\end{pmatrix*}$
b = 1
Number of SRs: 5
$\begin{pmatrix*}[l]\sqrt{3}\texttt{A}(\overline{\texttt{B}}_\texttt{d}^0 \to \texttt{D}^0 \Xi^{*-} \overline{\Sigma}^{*+}) + 2\texttt{A}(\overline{\texttt{B}}_\texttt{d}^0 \to \texttt{D}^0 \Sigma^{*-} \overline{\Delta}^{+}) + 2\texttt{A}(\overline{\texttt{B}}_\texttt{s}^0 \to \texttt{D}^0 \Xi^{*-} \overline{\Omega}^{+}) + \sqrt{3}\texttt{A}(\overline{\texttt{B}}_\texttt{s}^0 \to \texttt{D}^0 \Sigma^{*-} \overline{\Xi}^{*+}) \\\hfill\vdots\hfill\end{pmatrix*}$
b = 2
Number of SRs: 3
$\begin{pmatrix*}[l]-\sqrt{3}\texttt{A}(\overline{\texttt{B}}_\texttt{d}^0 \to \texttt{D}^0 \Xi^{*-} \overline{\Sigma}^{*+}) - \texttt{A}(\overline{\texttt{B}}_\texttt{d}^0 \to \texttt{D}^0 \Sigma^{*-} \overline{\Delta}^{+}) - \texttt{A}(\overline{\texttt{B}}_\texttt{d}^0 \to \texttt{D}^0 \Omega^{-} \overline{\Xi}^{*+}) + \texttt{A}(\overline{\texttt{B}}_\texttt{s}^0 \to \texttt{D}^0 \Delta^{-} \overline{\Sigma}^{*+}) \\\quad +\texttt{A}(\overline{\texttt{B}}_\texttt{s}^0 \to \texttt{D}^0 \Xi^{*-} \overline{\Omega}^{+}) + \sqrt{3}\texttt{A}(\overline{\texttt{B}}_\texttt{s}^0 \to \texttt{D}^0 \Sigma^{*-} \overline{\Xi}^{*+}) \\\hfill\vdots\hfill\end{pmatrix*}$
b = 3
Number of SRs: 1
$\begin{pmatrix*}[l]-\texttt{A}(\overline{\texttt{B}}_\texttt{d}^0 \to \texttt{D}^0 \Delta^{-} \overline{\Delta}^{+}) - 3\texttt{A}(\overline{\texttt{B}}_\texttt{d}^0 \to \texttt{D}^0 \Xi^{*-} \overline{\Xi}^{*+}) + 3\texttt{A}(\overline{\texttt{B}}_\texttt{d}^0 \to \texttt{D}^0 \Xi^{*-} \overline{\Sigma}^{*+}) + \sqrt{3}\texttt{A}(\overline{\texttt{B}}_\texttt{d}^0 \to \texttt{D}^0 \Sigma^{*-} \overline{\Delta}^{+}) \\\quad -3\texttt{A}(\overline{\texttt{B}}_\texttt{d}^0 \to \texttt{D}^0 \Sigma^{*-} \overline{\Sigma}^{*+}) - \texttt{A}(\overline{\texttt{B}}_\texttt{d}^0 \to \texttt{D}^0 \Omega^{-} \overline{\Omega}^{+}) + \sqrt{3}\texttt{A}(\overline{\texttt{B}}_\texttt{d}^0 \to \texttt{D}^0 \Omega^{-} \overline{\Xi}^{*+}) - \texttt{A}(\overline{\texttt{B}}_\texttt{s}^0 \to \texttt{D}^0 \Delta^{-} \overline{\Delta}^{+}) \\\quad +\sqrt{3}\texttt{A}(\overline{\texttt{B}}_\texttt{s}^0 \to \texttt{D}^0 \Delta^{-} \overline{\Sigma}^{*+}) + \sqrt{3}\texttt{A}(\overline{\texttt{B}}_\texttt{s}^0 \to \texttt{D}^0 \Xi^{*-} \overline{\Omega}^{+}) - 3\texttt{A}(\overline{\texttt{B}}_\texttt{s}^0 \to \texttt{D}^0 \Xi^{*-} \overline{\Xi}^{*+}) + 3\texttt{A}(\overline{\texttt{B}}_\texttt{s}^0 \to \texttt{D}^0 \Sigma^{*-} \overline{\Xi}^{*+}) \\\quad -3\texttt{A}(\overline{\texttt{B}}_\texttt{s}^0 \to \texttt{D}^0 \Sigma^{*-} \overline{\Sigma}^{*+}) - \texttt{A}(\overline{\texttt{B}}_\texttt{s}^0 \to \texttt{D}^0 \Omega^{-} \overline{\Omega}^{+})\end{pmatrix*}$
-------------------------
Amplitude-squared sum rules
b = 0
Number of SRs: 7
$\begin{pmatrix*}[l]\Gamma(\overline{\texttt{B}}_\texttt{s}^0 \to \texttt{D}^0 \Xi^{*-} \overline{\Omega}^{+}) - \Gamma(\overline{\texttt{B}}_\texttt{d}^0 \to \texttt{D}^0 \Sigma^{*-} \overline{\Delta}^{+})\\\hfill\vdots\hfill\end{pmatrix*}$
b = 1
Number of SRs: 4
$\begin{pmatrix*}[l]6\Gamma(\overline{\texttt{B}}_\texttt{d}^0 \to \texttt{D}^0 \Xi^{*-} \overline{\Xi}^{*+}) + 2\Gamma(\overline{\texttt{B}}_\texttt{d}^0 \to \texttt{D}^0 \Sigma^{*-} \overline{\Delta}^{+}) - 3\Gamma(\overline{\texttt{B}}_\texttt{d}^0 \to \texttt{D}^0 \Sigma^{*-} \overline{\Sigma}^{*+}) - 3\Gamma(\overline{\texttt{B}}_\texttt{d}^0 \to \texttt{D}^0 \Omega^{-} \overline{\Omega}^{+}) \\\quad -3\Gamma(\overline{\texttt{B}}_\texttt{s}^0 \to \texttt{D}^0 \Delta^{-} \overline{\Delta}^{+}) + 2\Gamma(\overline{\texttt{B}}_\texttt{s}^0 \to \texttt{D}^0 \Xi^{*-} \overline{\Omega}^{+}) - 3\Gamma(\overline{\texttt{B}}_\texttt{s}^0 \to \texttt{D}^0 \Xi^{*-} \overline{\Xi}^{*+}) + 6\Gamma(\overline{\texttt{B}}_\texttt{s}^0 \to \texttt{D}^0 \Sigma^{*-} \overline{\Sigma}^{*+}) \\\\ 4\Gamma(\overline{\texttt{B}}_\texttt{d}^0 \to \texttt{D}^0 \Xi^{*-} \overline{\Xi}^{*+}) + \Gamma(\overline{\texttt{B}}_\texttt{d}^0 \to \texttt{D}^0 \Xi^{*-} \overline{\Sigma}^{*+}) - 2\Gamma(\overline{\texttt{B}}_\texttt{d}^0 \to \texttt{D}^0 \Sigma^{*-} \overline{\Sigma}^{*+}) - 2\Gamma(\overline{\texttt{B}}_\texttt{d}^0 \to \texttt{D}^0 \Omega^{-} \overline{\Omega}^{+}) \\\quad -2\Gamma(\overline{\texttt{B}}_\texttt{s}^0 \to \texttt{D}^0 \Delta^{-} \overline{\Delta}^{+}) - 2\Gamma(\overline{\texttt{B}}_\texttt{s}^0 \to \texttt{D}^0 \Xi^{*-} \overline{\Xi}^{*+}) + \Gamma(\overline{\texttt{B}}_\texttt{s}^0 \to \texttt{D}^0 \Sigma^{*-} \overline{\Xi}^{*+}) + 4\Gamma(\overline{\texttt{B}}_\texttt{s}^0 \to \texttt{D}^0 \Sigma^{*-} \overline{\Sigma}^{*+}) \\\\ 6\Gamma(\overline{\texttt{B}}_\texttt{d}^0 \to \texttt{D}^0 \Xi^{*-} \overline{\Xi}^{*+}) - 3\Gamma(\overline{\texttt{B}}_\texttt{d}^0 \to \texttt{D}^0 \Sigma^{*-} \overline{\Sigma}^{*+}) - 3\Gamma(\overline{\texttt{B}}_\texttt{d}^0 \to \texttt{D}^0 \Omega^{-} \overline{\Omega}^{+}) + 2\Gamma(\overline{\texttt{B}}_\texttt{d}^0 \to \texttt{D}^0 \Omega^{-} \overline{\Xi}^{*+}) \\\quad -3\Gamma(\overline{\texttt{B}}_\texttt{s}^0 \to \texttt{D}^0 \Delta^{-} \overline{\Delta}^{+}) + 2\Gamma(\overline{\texttt{B}}_\texttt{s}^0 \to \texttt{D}^0 \Delta^{-} \overline{\Sigma}^{*+}) - 3\Gamma(\overline{\texttt{B}}_\texttt{s}^0 \to \texttt{D}^0 \Xi^{*-} \overline{\Xi}^{*+}) + 6\Gamma(\overline{\texttt{B}}_\texttt{s}^0 \to \texttt{D}^0 \Sigma^{*-} \overline{\Sigma}^{*+}) \\\\ \Gamma(\overline{\texttt{B}}_\texttt{d}^0 \to \texttt{D}^0 \Delta^{-} \overline{\Delta}^{+}) + 3\Gamma(\overline{\texttt{B}}_\texttt{d}^0 \to \texttt{D}^0 \Xi^{*-} \overline{\Xi}^{*+}) - 3\Gamma(\overline{\texttt{B}}_\texttt{d}^0 \to \texttt{D}^0 \Sigma^{*-} \overline{\Sigma}^{*+}) - \Gamma(\overline{\texttt{B}}_\texttt{d}^0 \to \texttt{D}^0 \Omega^{-} \overline{\Omega}^{+}) \\\quad -\Gamma(\overline{\texttt{B}}_\texttt{s}^0 \to \texttt{D}^0 \Delta^{-} \overline{\Delta}^{+}) - 3\Gamma(\overline{\texttt{B}}_\texttt{s}^0 \to \texttt{D}^0 \Xi^{*-} \overline{\Xi}^{*+}) + 3\Gamma(\overline{\texttt{B}}_\texttt{s}^0 \to \texttt{D}^0 \Sigma^{*-} \overline{\Sigma}^{*+}) + \Gamma(\overline{\texttt{B}}_\texttt{s}^0 \to \texttt{D}^0 \Omega^{-} \overline{\Omega}^{+})\end{pmatrix*}$
b = 2
Number of SRs: 0
No sum rules found at this order.
b = 3
Number of SRs: 0
No sum rules found at this order.
\end{Verbatim}

\subsection{Mass sum rules for $\Omega_B^-$}
\label{sec:mass sum rules}

Finally, we use \verb|FlaSR| to derive mass sum rules, by which we mean sum rules between hadron masses, for the four hadrons in the $\Omega_B^-$ multiplet defined in Eq.~\eqref{eq:BDOOmultiplet}. Since the properly normalized diagonal matrix element of the QCD Hamiltonian between one-hadron states at rest gives the corresponding hadron mass, the amplitude sum rules for the system $\Omega_B^- \to \Omega_B^-$ can be interpreted as mass sum rules, i.e. as sum rules between the masses of the members of the $\Omega_B^-$ multiplet.

We demonstrate how to use \verb|FlaSR| to derive the mass sum rules for the $\Omega_B^-$ multiplet, express the amplitudes in terms of symbols that represent mass, and index amplitudes by particle names. We begin by defining the multiplets in the $\Omega_B^- \to \Omega_B^-$ system, which has a singlet Hamiltonian. As a reminder, in the physical mode we input the coefficients of the Hamiltonian, which for a singlet is simply \verb|{1}|.
\begin{Verbatim}[frame=single,commandchars=\\\{\},codes={\catcode`$=3\catcode`^=7\catcode`_=8},breaklines=true]
OBm = \{"$\Delta^-$","$\Sigma^{*-}$","$\Xi^{*-}$","$\Omega^{-}$"\};
H = \{1\};
\end{Verbatim}

We generate the sum rules for the $\Omega_B^- \to \Omega_B^-$ system in the physical input mode (\verb|phys -> True|):
\begin{Verbatim}[frame=single,commandchars=\\\{\},codes={\catcode`$=3\catcode`^=7\catcode`_=8},breaklines=true]
system = generateSRs[\{OBm\}, \{H\}, \{OBm\}, phys -> True];
\end{Verbatim}

To view only the amplitude table section of the output, we can run a \verb|FlaSR| function called \verb|printAmps|:
\begin{Verbatim}[frame=single,commandchars=\\\{\},codes={\catcode`$=3\catcode`^=7\catcode`_=8},breaklines=true]
printAmps[system];
\end{Verbatim}

The $\Omega_B^- \to \Omega_B^-$ system has the following amplitude table:
\begin{Verbatim}[frame=single,commandchars=\\\{\},codes={\catcode`$=3\catcode`^=7\catcode`_=8\catcode`&=4}]
Amplitude table
Number of amplitudes: 4
a/s definitions: $\texttt{a}_\texttt{i} = \texttt{A}_\texttt{i} - \overline{\texttt{A}}_\texttt{i}$, $\texttt{s}_\texttt{i} = \texttt{A}_\texttt{i} + \overline{\texttt{A}}_\texttt{i}$
$\Delta$/$\Sigma$ definitions: $\Delta_\texttt{i} = |\texttt{A}_\texttt{i}|^2 - |\overline{\texttt{A}}_\texttt{i}|^2$, $\Sigma_\texttt{i} = |\texttt{A}_\texttt{i}|^2 + |\overline{\texttt{A}}_\texttt{i}|^2$
$\begin{matrix*}[l]\texttt{Process} && \texttt{QN label} && \texttt{n-tuple} && \texttt{Coord} && \texttt{CKM}\\\Delta^- \to \Delta^- && +\frac{3}{2} \overset{0}{\to} +\frac{3}{2} && \texttt{(---,+++)} && \texttt{(1,1)} && 1\\\Omega^- \to \Omega^- && -\frac{3}{2} \overset{0}{\to} -\frac{3}{2} && \texttt{(+++,---)} && && 1\\\\\Sigma^{*-} \to \Sigma^{*-} && +\frac{1}{2} \overset{0}{\to} +\frac{1}{2} && \texttt{(--+,-++)} && \texttt{(1,2)} && 1\\\Xi^{*-} \to \Xi^{*-} && -\frac{1}{2} \overset{0}{\to} -\frac{1}{2} && \texttt{(-++,--+)} && && 1\end{matrix*}$
\end{Verbatim}

By default, the amplitudes will be indexed by their processes. To index amplitudes by particle names rather than process names (e.g., to use $\Delta^-$ rather than $\Delta^- \to \Delta^-$), we use
\begin{center}
    \texttt{labelAmps[system, colName, labels, \ldots]}
\end{center}
to add a column of custom labels to the amplitude table. This function takes in the \verb|system| association, a string \verb|colName| assigning a name to the column, and a list \verb|labels| of labels for the amplitudes in the order they appear in the amplitude table. The default labeling mode is for individual amplitudes, though there is also an option for labeling amplitude pairs; see \href{https://github.com/Flavor-Sum-Rules/FlaSR}{GitHub} for details.

In \verb|FlaSR|, we add a column called \verb|"Particle"| to the amplitude table with the names of the particles in the $\Omega_B^-$ multiplet as labels. Note that the labels follow the order of the amplitude table, not the multiplet:
\begin{Verbatim}[frame=single,commandchars=\\\{\},codes={\catcode`$=3\catcode`^=7\catcode`_=8},breaklines=true]
labelAmps[system, "Particle", \{"$\Delta^-$", "$\Omega^-$", "$\Sigma^{*-}$", "$\Xi^{*-}$"\}];
\end{Verbatim}

For the formatting options for \verb|printSystem|, we use the symbol \verb|m| instead of \verb|A| to represent amplitudes (\verb|ampType -> {m}|). We subscript the amplitudes using our custom labels (\verb|ampFormat -> "Particle"|) and suppress amplitudes-squared from being displayed since the amplitude-level relations already give observable relations (\verb|showA2SRs -> False|):
\begin{Verbatim}[frame=single,commandchars=\\\{\},codes={\catcode`$=3\catcode`^=7\catcode`_=8},breaklines=true]
printSystem[system, ampType -> \{m\}, showA2SRs -> False, ampFormat -> "Particle"];
\end{Verbatim}

The full output is given below. Notice that the column of particle names now appears in the amplitude table:
\begin{Verbatim}[frame=single,commandchars=\\\{\},codes={\catcode`$=3\catcode`^=7\catcode`_=8\catcode`&=4}]
System: \{0,$\frac{3}{2}$,$\frac{3}{2}$\}
-------------------------
Number of would-be doublets: 6
In: \{$\frac{3}{2}$\}
H: \{0\}
Out: \{$\frac{3}{2}$\}
-------------------------
Amplitude table
Number of amplitudes: 4
a/s definitions: $\texttt{a}_\texttt{i} = \texttt{A}_\texttt{i} - \overline{\texttt{A}}_\texttt{i}$, $\texttt{s}_\texttt{i} = \texttt{A}_\texttt{i} + \overline{\texttt{A}}_\texttt{i}$
$\Delta$/$\Sigma$ definitions: $\Delta_\texttt{i} = |\texttt{A}_\texttt{i}|^2 - |\overline{\texttt{A}}_\texttt{i}|^2$, $\Sigma_\texttt{i} = |\texttt{A}_\texttt{i}|^2 + |\overline{\texttt{A}}_\texttt{i}|^2$
$\begin{matrix*}[l]\texttt{Particle} && \texttt{Process} && \texttt{QN label} && \texttt{n-tuple} && \texttt{Coord} && \texttt{CKM}\\\Delta^- && \Delta^- \to \Delta^- && +\frac{3}{2} \overset{0}{\to} +\frac{3}{2} && \texttt{(---,+++)} && \texttt{(1,1)} && 1\\\Omega^- && \Omega^- \to \Omega^- && -\frac{3}{2} \overset{0}{\to} -\frac{3}{2} && \texttt{(+++,---)} && && 1\\\\\Sigma^{*-} && \Sigma^{*-} \to \Sigma^{*-} && +\frac{1}{2} \overset{0}{\to} +\frac{1}{2} && \texttt{(--+,-++)} && \texttt{(1,2)} && 1\\\Xi^{*-} && \Xi^{*-} \to \Xi^{*-} && -\frac{1}{2} \overset{0}{\to} -\frac{1}{2} && \texttt{(-++,--+)} && && 1\end{matrix*}$
-------------------------
Amplitude sum rules
b = 0
Number of SRs: 2
$\begin{pmatrix}\texttt{m}_{\Delta^-} & \texttt{m}_{\Omega^-} & \texttt{m}_{\Sigma^{*-}} & \texttt{m}_{\Xi^{*-}}\\\hline 1 & -1 & 0 & 0\\0 & 0 & 1 & -1\end{pmatrix}$
b = 1
Number of SRs: 1
$\begin{pmatrix}\texttt{m}_{\Delta^-} & \texttt{m}_{\Omega^-} & \texttt{m}_{\Sigma^{*-}} & \texttt{m}_{\Xi^{*-}}\\\hline 1 & 1 & -1 & -1\end{pmatrix}$
b = 2
Number of SRs: 1
$\begin{pmatrix}\texttt{m}_{\Delta^-} & \texttt{m}_{\Omega^-} & \texttt{m}_{\Sigma^{*-}} & \texttt{m}_{\Xi^{*-}}\\\hline 1 & -1 & -3 & 3\end{pmatrix}$
\end{Verbatim}

We read off the following mass sum rules from the \verb|FlaSR| output above:
\begin{equation}
m_{\Delta^-}-m_{\Omega^-}=\mathcal{O}(\varepsilon)\,, \qquad
m_{\Sigma^{*-}}-m_{\Xi^{*-}}=\mathcal{O}(\varepsilon)\,,
\end{equation}
\begin{equation}
m_{\Delta^-}+m_{\Omega^-}-m_{\Sigma^{*-}}-m_{\Xi^{*-}}
=\mathcal{O}(\varepsilon^2)\,,
\end{equation}
\begin{equation}
m_{\Delta^-}-m_{\Omega^-}-3m_{\Sigma^{*-}}+3m_{\Xi^{*-}}
=\mathcal{O}(\varepsilon^3)\,.
\end{equation}

\section{Conclusion}
\label{sec:conclusion}

\texttt{FlaSR} is a Mathematica package for the automatic generation of $SU(2)$ flavor sum rules, with a primary emphasis on $U$-spin. Given the irreducible representations of the initial and final states and of the symmetry-limit Hamiltonian, the package generates the corresponding system of processes and computes its amplitude and amplitude-squared sum rules. The package is based on recent insights into the structure of flavor sum rules that allow the sum rules to be derived without performing an explicit Clebsch--Gordan decomposition. Our algorithm guarantees the derivation of the complete set of amplitude sum rules at every order in the symmetry-breaking expansion at which such relations exist, but it does not provide the same guarantee for amplitude-squared sum rules.

We illustrated the use of \texttt{FlaSR} on several representative examples, including two- and three-body hadronic decays, a system with identical final-state multiplets, and mass sum rules. These examples demonstrate both the basic workflow of the package and the range of systems to which the formalism can be applied. Natural future directions include generalizing the procedure to derive the complete set of rate sum rules for a given system, extending the algorithm to other $SU(2)$ flavor subgroups in cases where the completeness condition stated in Section~\ref{sec:scope} is not satisfied, such as in isospin- and $V$-spin-related hadronic weak decays, and implementing direct-sum systems in the program.

We expect \texttt{FlaSR} to be useful as a practical tool for discovering new systems with higher-order sum rules. Such relations can provide concrete targets for phenomenological and experimental studies. In particular, systematic searches for $U$-spin systems supporting higher-order rate sum rules and their tests might be able to shed light on non-perturbative QCD effects, test the convergence of the flavor-breaking expansion, and sharpen the interpretation of possible deviations from Standard Model expectations. We encourage the use of \texttt{FlaSR} in mapping the landscape of systems with higher order sum rules and in developing the corresponding phenomenological analyses.

\section*{Acknowledgments}
We would like to thank Yuval Grossman, Guglielmo Papiri, and Stefan Schacht for related collaborations, for testing the package, and for providing valuable feedback. We are grateful to Guglielmo Papiri for suggesting the ``squaring'' procedure implemented in \texttt{FlaSR}. This material is based upon work supported by the U.S. Department of Energy, Office of Science, Office of High Energy Physics, under Award Number DE-SC0011632.




\bibliographystyle{elsarticle-num}
\bibliography{refs}






\end{document}